\documentclass[fleqn,usenatbib]{mnras}

 \usepackage{newtxtext,newtxmath}
\usepackage[T1]{fontenc}

\DeclareRobustCommand{\VAN}[3]{#2}
\let\VANthebibliography\thebibliography
\def\thebibliography{\DeclareRobustCommand{\VAN}[3]{##3}\VANthebibliography}

\newcommand{\CI}{\mbox{C\,{\sc i}}}

\usepackage{graphicx}	% Including figure files
\usepackage{amsmath}	% Advanced maths commands
\usepackage{multirow}

\title[Metal Factories in the Early Universe]{Metal Factories in the Early Universe}

\author[S.A. Eales et al.]{
Stephen Eales,$^{1}$\thanks{E-mail: steve.a.eales@gmail.com}
Haley Gomez,$^{1}$
Loretta Dunne,$^{1}$
Simon Dye,$^{2}$
and Matthew W. L. Smith$^{1}$
\\
$^{1}$Cardiff Hub for Astrophysics Research and Technology, Cardiff University, The Parade, Cardiff CF24 3AA, UK\\
$^{2}$School of Physics and Astronomy, University of Nottingham, University Park, Nottingham NG7 2RD, UK\\
}

\date{Accepted XXX. Received YYY; in original form ZZZ}

\pubyear{2015}

\begin{document}
\label{firstpage}
\pagerange{\pageref{firstpage}--\pageref{lastpage}}
\maketitle

\begin{abstract}
We have estimated the mass of metals in the molecular gas in 13 dusty star-forming galaxies at $z \sim 4$ in which the
gas, based on previous observations, lies in a cold rotating disk.
We estimated the metal masses
using either the submillimetre line or continuum emission
from three tracers of the overall metal content -
carbon atoms, carbon monoxide molecules and dust grains -
using the first 
simultaneous calibration of all three tracers. 
We obtain very similar mass
estimates from the different tracers, which are
similar to the entire metal content of
a present-day massive early-type galaxy.
We used the dynamical masses of these galaxies to estimate an upper limit on the mass of the molecular gas in each
galaxy, allowing us to estimate a lower limit on the metal abundance of the
gas, finding values for many of the galaxies well above the solar value.
We show 
that the high metal masses and metal 
abundances are what is expected shortly after the formation of
a galaxy for a top-heavy IMF.
We suggest a scenario for galaxy evolution in which massive galaxies reach a
high metal abundance during their formation phase, which is then
gradually reduced by dry mergers with lower mass galaxies. 
We show that the metals in the outflows 
from high-redshift dusty star-forming galaxies can quantitatively explain the
long-standing puzzle that such a large fraction of the metals 
in galaxy clusters ($\simeq$0.75) is in the intracluster
gas rather than in the galaxies themselves. 

\end{abstract}


%%%%%%%%%%%%%%%%%%%%%%%%%%%%%%%%%%%%%%%%%%%%%%%%%%
 
%%%%%%%%%%%%%%%%% BODY OF PAPER %%%%%%%%%%%%%%%%%%

\section{Introduction}

One of the most distinctive properties of the galaxy population today is the strong relationship
between metal abundance and stellar mass, with metal abundance increasing with
stellar mass and
reaching a value of about twice the solar value at the highest masses
\citep{gallazzi2005}. 
Prior to the launch of the James Webb Space Telescope (JWST), there was 
evidence that this relationship already existed
for galaxies at $z \sim 3-5$, although with the metal abundances 
a factor of $\simeq$4 lower than for galaxies with the same stellar mass in the
universe today \citep{cullen2019}. 
Recent JWST observations have now shown this relationship was already in place
at $z \sim 10$, with the metal abundances lower by a factor of $~4-10$ at the same
stellar mass from the relationship today
\citep{langeroodi2023,heintz2023,fujimoto2023,curti2023}.

Lower metal abundances for galaxies at high redshifts are, of
course, not really surprising
because there has been less time then
for metals to have been made in stars and ejected into the ISM.
Nevertheless, there are signs in the universe today that there
must have been a period of rapid metal production early in the
history of the universe.
The best evidence comes from rich clusters of galaxies, in which 
the strong gravitational
fields ensure that no metals should have been lost from the cluster.
In nearby rich clusters, roughly 75\,per\,cent of the metals are in the
intracluster gas rather than in the 
galaxies \citep{portinari2004,loewenstein2013,renzini2014,mernier2021}. 
Clusters
today are dominated by elliptical and lenticular galaxies - early-type
galaxies - in which the current rate of star formation, and therefore
metal production, is low. The best estimates, based on the spectra
of these galaxies, suggest that most of their stars were formed during
the first two billion years after the big bang during a burst of star formation  lasting only
$\simeq5 \times 10^8$
years 
\citep{thomas2010}. It 
therefore
seems likely that most of the metals that are in the intracluster gas
today were formed during this early period of star formation and then ejected
from the galaxies, although models based on this assumption have so far failed
to reproduce the mass of metals in the intracluster gas by a
factor of between 2 and 10 \citep{loewenstein2013,renzini2014}.

The estimated star-formation rates of these galaxies during this early period of star formation 
lie in the range 100-1000 
$\rm M_{\odot}\ year^{-1}$. 
Twenty-five years after they were discovered, the galaxies 
found
in the first submillimetre surveys \citep{smail1997,hughes1998,barger1998,eales1999} 
are the still the only galaxies in the correct redshift range and with
the necessary star-formation rates to be the ancestors of the massive early-type galaxies in present-day clusters
\citep{hughes1998,lilly1999}.
The wide-field submillimetre surveys with the {\it Herschel Space
Observatory} and the South Pole Telescope (SPT) have discovered many
examples of these dusty star-forming galaxies (henceforth DSFGs) with a star-formation rate $\rm >1000\ M_{\odot}\ year^{-1}$
\citep{bakx2018,reuter2020}.

The DSFGs must contain a large mass of dust, which is mostly
composed of metals, and they are therefore an exception
to the
metal-poor high-redshift universe revealed by the early
JWST observations. 
The dust, however, makes it challenging to use standard optical techniques
to measure their metal abundance, with estimates of the visual
extinction often having
extreme values (e.g. $A_V > 450$ - \citet{harrington2021}). On top 
of this problem, there
is the additional problem that many of the brightest DSFGs are gravitationally
lensed, which means there is a bright galaxy sitting in front of the
DSFG, contaminating its optical emission.

In this paper, we have developed an alternative way of estimating the
mass of metals and the metal abundance in these galaxies. 
A widely used technique for estimating the mass of the molecular gas in a 
galaxy is to use 
the luminosity of a tracer of the gas to estimate its mass.
Carbon monoxide 
molecules \citep{bolatto2013,tacconi2020}, carbon atoms \citep{papadopoulos2004}, and 
dust grains \citep{eales2012,scoville2016,scoville2017,tacconi2020} have all been tried.

Whichever of the tracers is used, the calibration of this method is ultimately based almost always on observations of the
ISM in the Galaxy, which suggests two obvious objections.
The first is that an implicit assumption of the method is that the conditions in
the galaxy of interest - the density, temperature and structure of its molecular gas, the physical and chemical properties of its
dust grains - are the same as in the Galaxy. The second objection,
since all the tracers are made of metals, is that an implicit assumption of
the method is that the metal abundance is
the same as in the Galaxy. Given the evolution seen in
the relationship between metal abundance and stellar mass (see above), this assumption seems particularly 
dubious if the galaxy of interest is at high redshift
\citep{scoville2016,scoville2017,tacconi2020}.

The second objection can be
avoided if the method, as we do here, is used to estimate the mass
of the metals rather than the
mass of the molecular gas, since it then 
does not rely on any assumption about the
metal abundance. Dunne et al. (2022) have made
the first attempt to calibrate the tracer method which treats
all tracers equally rather than relying on one as a gold standard
to calibrate the others (Appendix A). We have used these
new calibration factors to estimate the mass of metals in the
molecular gas for a sample of DSFGs at $z \sim 4$.
Although our method avoids the assumption of a
universal metal abundance, it still relies on the assumption that
the critical properties of the ISM on which the emission from the tracers depend
are the same as in the Galactic ISM.

This method is a way of estimating the total mass of metals
but not the metal abundance, which {\it does} require knowledge of the mass
of the molecular gas. However, we can estimate an upper limit on the mass of the
gas from an estimate of the dynamical mass of the galaxy. 
A large proportion of the extreme DSFGs discovered
with {\it Herschel} and the South Pole Telescope are magnified by gravitational
lenses,
which has made it possible to
investigate the motion of the gas with resolution as high as 50 parcsec
\citep{dye2015}. Observations of the gas kinematics have revealed
that the gas in many of these
galaxies is distributed in a cold rotating
disk \citep{hodge2012,dye2015,rizzo2020,
neeleman2020,fraternali2021,rizzo2021,lelli2021,dye2022}, which makes it 
possible to estimate the mass of the
galaxy from the circular velocity of the gas.
By estimating the dynamical mass of each galaxy, we have estimated an upper
limit on the mass of the molecular gas, which 
has allowed us to estimate a
lower limit on the metal abundance. 

The arrangement of the paper is as follows. \S 2 describes our sample
of galaxies. In \S 3 we describe our method for estimating 
the mass of metals
and the application of this method to our sample.
In \S 4, we use the dynamical masses
of the galaxies to estimate an upper limit on the mass of the molecular gas in each galaxy and thus
a lower limit on the galaxy's
metal abundance. In this section, we also propose
a solution to the well known paradox that the estimate of the mass of the
molecular gas in a high-redshift galaxy is often greater than the estimate of
its dynamical mass.
In \S 5, we use chemical evolution models to interpret
our results.
In \S 6
we discuss our results and we list our conclusions in \S 7.
We used the cosmological parameters given in \citet{planck2015}.

\section{The Sample}

We selected a  sample of 13 high-redshift galaxies for which there is evidence from the velocity profile (velocity versus radius) and from the high value of the ratio of rotational
velocity to velocity dispersion (\S 4.1)
that the gas in the galaxy lies in a cold rotating disk (Table~\ref{Sample Table}).

All the galaxies 
also have observations that make it possible to estimate the mass of metals in the molecular gas, having at least
one, and in most cases more than one, of the following observations: spectral line observations of \CI 1-0; spectral line observations of either CO 1-0 or 2-1; continuum observations of
dust. 

The CO observations were mostly made with the 
Australia Telescope Compact Array \citep{aravena2016}, which have
a resolution of $\rm \simeq5\ arcsec$. The \CI\ observations were
all made with the Atacama Millimeter Array (ALMA) \citep{bothwell2017,dye2022}. Apart from the high-resolution
observations of ID141 \citep{dye2022}, these have a resolution
of $\rm \simeq 5\ arcsec$ \citep{bothwell2017}. Our main sources of
dust observations were the compilation of multi-frequency global
dust measurements for the SPT galaxies \citep{reuter2020} and
our reanalysis of data in the ALMA archive (Section 3.1). In the
case of the dust, there is almost always a high-resolution 
continuum image which makes it possible to be confident that
the dust is within the region covered by the dynamical analysis
e.g. \citet{dye2015,dye2022,spilker2016}.

Eight of the 13 galaxies are gravitationally lensed, and for each of these there is a lensing model with an estimate of the magnification and
its uncertainty (Table~\ref{Sample Table}). Twelve of the sample are at $z>4$ when the universe was less than 1.6 billion years old.  

\begin{table}
        \caption{The Sample}
        \label{Sample Table}
        \begin{tabular}{lccc} % four columns, alignment for each
                \hline
                \multicolumn{1}{c}{\multirow{2}{*}{Source}} & \multicolumn{1}{c}{\multirow{2}{*}{Redshift}} & Gravitational & \multicolumn{1}{c}{\multirow{2}{*}{Reference}}\\
                 &  & Magnification & \\
                \hline
                SPT0113-46 & 4.233 & 23.86\,$\pm$\,0.51 & 1 \\
                ALESS073.1 & 4.76\phantom{0} &  & 2 \\
                SPT0345-47 & 4.296 & \phantom{0}7.95\,$\pm$\,0.48 & 1 \\
                SPT0418-47 & 4.225 & 32.7\phantom{0}\,$\pm$\,2.66 & 1 \\
                SPT0441-46 & 4.48\phantom{0} & 12.73\,$\pm$\,0.96 & 1 \\
                DLA0817g & 4.26\phantom{0} &  & 3 \\
                SDP81 & 3.042 & 15.9\phantom{0}\,$\pm$\,0.7\phantom{0} & 4 \\
                AZTEC/C159 & 4.57\phantom{0} & & 5,6 \\
                J1000+0234& 4.54\phantom{0} &  & 5,6 \\
                GN20 & 4.05\phantom{0} &  & 7 \\
                ID141 & 4.24\phantom{0} & 5.5$^a$ & 8  \\
                SPT2132-58 & 4.768 & \phantom{0}5.72\,$\pm$\,0.54 & 1 \\
                SPT2146-55 & 4.567 & \phantom{0}6.65\,$\pm$\,0.41 & 1 \\
                \hline
        \end{tabular}
\begin{flushleft}
Notes: a - \citet{dye2022} do not give an error for 
the gravitational magnification factor, but all the gas
and metal masses in this paper have been calculated from 
demagnified fluxes which have errors that incorporate
the uncertainties in the lensing model. Also note that the ISM masses given 
by \citet{dye2022} do not
include a correction for the effect of the cosmic microwave background (\S 3.2).

References: 1 - \citet{reuter2020}; 2 - \citet{lelli2021}; 3 - \citet{neeleman2020}; 4 - \citet{dye2015};
5 - \citet{jones2017}, 6 - \citet{fraternali2021}; 7 - \citet{hodge2012}; 8 - \citet{dye2022}.
\end{flushleft}
\end{table}

\section{Metal Masses}

\subsection{The Method}

Since molecular hydrogen does 
not itself emit spectral lines at the 
typical temperature of the molecular phase of the ISM, the standard method to estimate
the mass of the molecular gas is from the luminosity ($L$) of some
`tracer'.
The mass of the molecular gas is then given by the
equation:

\begin{equation}
M_{\rm mol} = \alpha L
\end{equation}

\noindent in which $\alpha$ is a calibration factor.
Carbon atoms (\CI ), CO and dust grains have all been used as tracers.
The calibration factor
is almost always ultimately based 
on observations of the molecular gas in the Galaxy, from which it is 
possible to estimate the gas mass more directly, for
example from gamma-ray observations or from the dynamical masses of molecular 
clouds \citep{bolatto2013,tacconi2020}. Over the last decade the application of this technique to high-redshift galaxies
has led to the conclusion that as much as 90\,per\,cent of the 
baryonic mass in a high-redshift galaxy is in the form of gas
\citep{scoville2016,scoville2017,tacconi2018}.

As we noted above (Section 1), an obvious concern with
this technique is that it relies on the assumption that the metal
abundance is everywhere the same, which is especially dangerous
when the galaxy is at high redshift.
Nevertheless, although sometimes an attempt is 
made to correct the value of $\alpha$ 
for this effect
using a measurement of the metal 
abundance in the galaxy \citep{tacconi2020},  
this measurement rarely has enough precision since 
the standard optical techniques for estimating metal abundance
all themselves have large systematic uncertainties
\citep{maiolino2019}. Therefore, for the want 
of anything better, the assumption that the
calibration factors have the same values as in the Galaxy is 
the one that is often made.

In this paper, we try to avoid this problem by using observations of the tracers to estimate the mass of the metals
rather than the mass of the molecular gas. We calculate the mass of metals in a galaxy from the luminosity
of a tracer, $L$ using the equation:

\begin{equation}
M_{} = \alpha_{\rm MW}\, A_{\rm MW}\, L
\end{equation}

\noindent In this equation, $\rm \alpha_{\rm MW}$ is the 
same calibration factor as in equation (1), to which we have added
a subscript to show it is
ultimately based on observations in the Galaxy
(Milky Way), and $A_{\rm MW}$ is the 
metal
abundance in the ISM in the Galaxy. We assume a metal abundance in the ISM in the Galaxy of 86.9, based on the latest
estimates of the gas-to-dust ratio (167.2) and the fraction of the metals that is incorporated in the
dust (0.52) \citep{romanduval}.

\begin{table}
        \caption{Continuum Measurements from Observations in the ALMA Archive}
        \label{tab:alma_table}
        \centering
        \begin{tabular}{lccc} % four columns, alignment for each
                \hline
                \multicolumn{1}{c}{\multirow{2}{*}{Source}} & Frequency & Flux density & ALMA\\
                & (GHz) & ($\mu$Jy)& project code\\
                \hline
                ALESS 073.1 & 92.74\phantom{0} & 136\,$\pm$\,22 & 2015.1.00040.S\\
                J1000+0234 & 103.345 & 180\,$\pm$\,31 & 2016.1.00171.S\\
                SDP81 & 88.850 & 676\,$\pm$\,98 & 2017.1.01694.S\\
                \hline
        \end{tabular}
\end{table}

This equation should still be correct even
if the metal abundance in the galaxy that
is being observed, $A_{\rm gal}$, is different from 
the value in the Galaxy. 
For the optically-thin tracers, dust and \CI\, \citep{harrington2021,padelis2022},
the value of the calibration factor for the galaxy, 
$\alpha_{\rm gal}$,
will be inversely proportional to the galaxy's metal abundance, 
and therefore $\alpha_{\rm gal} A_{\rm gal} = \alpha_{\rm MW} 
A_{\rm MW}$. This is not 
so obviously true for CO because the 
line emission is optically thick. Nevertheless, although
the dependence
of CO emission on metal abundance is still 
uncertain, the data does suggest a similar
relationship between $\alpha_{\rm gal}$ and $A_{\rm gal}$ (Figure 9 of
\citet{bolatto2013}).
We emphasise that although we are now not making the
assumption that the metal abundance is the same as in the Galaxy, we still need to make the assumption that the properties
of the gas and dust on which the calibration
factors depend are the same as in the Galaxy. For example,
when using dust as the tracer we implicitly assume that the dust
mass-opacity coefficient is the same as the Galactic value.

We estimated the metal masses for the galaxies in the sample 
using observations of the following tracers: carbon atoms (\CI ), carbon monoxide molecules (CO) and dust
grains, with calibration factors $\rm \alpha_{CI}$, $\rm \alpha_{\rm CO}$ and $\rm \alpha_{\rm 850\mu m}$, respectively.
We used the calibration factors from \citet{dunne2022}, which was the first attempt to calibrate all
three calibration factors simultaneously. The values we assume are
$\rm 1/\alpha_{850 \mu m} = 6.9\times10^{12}\ W\ Hz^{-1}\ M_{\odot}^{-1}$ for the dust,
$\rm \alpha_{\rm CO} = 4.0\ M_{\odot} (K\ km\ s^{-1}\ pc^2)^{-1}$ for the CO $J=1-0$ line, 
and 
$\rm \alpha_{CI}= 17\ M_{\odot} (K\ km\ s^{-1}\ pc^2)^{-1}$ for the CI $J=1-0$ line.
We assume the uncertainties in the calibration factors $\rm \alpha_{850 \mu m}$, $\alpha_{\rm CO}$
and $\rm \alpha_{CI}$ are 31\,per\,cent, 39\,per\,cent and 19\,per\,cent respectively, based on the values 
estimated in their paper.

We restricted ourselves to observations that are least likely to be affected by 
systematic uncertainties. We therefore used observations of atomic carbon in the \CI~ $J=1-0$ line 
but not in the \CI~ $J=2-1$  line
because of the recent evidence of subthermal excitation, which leads to 
large uncertainties in the estimates of the mass of molecular
gas from this line \citep{padelis2022}.  
We used 
observations of carbon monoxide in the $J=1-0$ and $J=2-1$ lines but not in the 
higher $J$ lines because the ratio of the line flux for these lines to the line flux for 
the $J=1-0$ line, on which the calibration is based, depends strongly on the temperature 
of the gas. We estimated a line flux for the $J=1-0$ line 
from the line flux in the $J=2-1$ line using the flux ratio 
$\rm CO 2-1/1-0 = 2.97\pm0.61$, 
which we derived from a study of carbon monoxide in a 
large sample of ultraluminous infrared galaxies \citep{padelis2012}. 

We only used dust as a tracer if there was an observation close in rest-frame wavelength to 
850 $\rm \mu m$, the wavelength at which the dust method is 
calibrated \citep{dunne2022}, which in practice mostly meant observations 
in ALMA Band 3. We found unpublished observations in the ALMA archive for three galaxies, and
we measured new flux densities for these, which are given in Table~\ref{tab:alma_table}.
Where there were continuum observations at many wavelengths \citep{reuter2020}
we fitted a modified blackbody
($F_\nu \propto B_{\nu} \nu^{\beta}$) to the flux densities
at wavelengths $\rm \ge 500 \mu m$, using a single dust temperature and 
a dust emissivity index $\rm \beta=2$. The advantage of estimating the 
dust temperature from only long-wavelength flux measurements is that the estimate should be 
closer to the mass-weighted dust temperature; the inclusion of flux measurements on the 
short-wavelength side of the blackbody peak biases the estimate towards the 
luminosity-weighted dust temperature \citep{eales1989}. 
Our estimated dust temperatures for these galaxies 
are listed in Table~\ref{tab:Metal Table} and they
support the argument that the mass-weighted dust
temperature, even for galaxies with very high star-formation rates,
is$\simeq$ 25 K \citep{scoville2016}. For these galaxies, we
have estimated the rest-frame flux density at 850 $\mu$m from
the best-fitting modified blackbody. For the galaxies for which there is only
a flux measurement at a single wavelength, we estimated
the rest-frame flux density at 850 $\mu$m
from this measurement and a modified blackbody with
a dust temperature of 25 K and $\beta = 2$.

Where necessary we corrected the line and continuum fluxes for the gravitational
magnification factors given in Table~\ref{Sample Table}.
We have estimated errors on our metal masses by adding the following errors in
quadrature: the error on the calibration factor (see above), the error on the
flux of the tracer and, where applicable, the errors in the gravitational magnification
factor and in the CO $2-1/1-0$ line ratio.
All the galaxies in the sample have observations of at least one tracer, 10 have 
observations of two tracers, and seven have observations of all three. 

\subsection{Corrections for the Cosmic Microwave Background}

At these redshifts, the cosmic microwave background (CMB) can have a large effect on submillimetre 
and radio observations, leading to underestimates in both line and 
continuum flux measurements. We have made corrections to the line and continuum fluxes using the method in \cite{dacunha2013}.
In this work we estimate metal masses both with and without a correction for the CMB
because of the potential for errors introduced by the assumptions necessary in making
this correction.

For optically thin radiation, which is the case for the dust continuum radiation at long wavelengths and the \CI~ $J=1-0$ line \citep{harrington2021,padelis2022}, 
the CMB corrections only depend on the temperature of the dust and the excitation temperature 
of the gas, respectively. For the dust, we corrected the continuum flux
densities for each galaxy using equations 12 and 18 in
\citet{dacunha2013}
on the assumption that the temperature
of the dust if it were not for the CMB would be 25 K (\S 3.1). We then estimated the rest-frame
850-$\mu$m flux density from the corrected flux densities using
the procedure described in \S 3.1.
The excitation temperature of the \CI~  $1-0$ line for DSFGs is $\simeq$25 K
\citep{padelis2022}, very similar to our estimates of the dust temperature (\S 3.1), 
which is
what one would expect if the dust temperature and kinetic temperature are the
same and the gas is in local thermodynamical equilibrium. 
For the CI, we therefore assumed the same temperature and used the same
equations to make the CMB correction.

The corrections for the CO $J=1-0$ and $J=2-1$ lines
are not so obvious
since the lines are optically thick, which 
means the corrections should depend on the opacity of the gas as well as 
on its temperature \citep{dacunha2013}. The uncertainties in the corrections from the 
inclusion of opacity are so large that we decided not to try to include the effect of
opacity but 
to apply the same method as for the optically-thin \CI~line.
Our 
CMB corrections for the CO lines therefore show the 
possible size of the effect of the CMB, but the actual size of the 
correction for these lines is very uncertain.

\begin{table*}
        \caption{The Mass of Metals }
        \label{tab:Metal Table}
        \begin{tabular}{l|ccccccccc} 
                \hline
 &   \multicolumn{3}{c}{Metal masses with no CMB} & \multicolumn{3}{c}{Metal masses with CMB} &
 & \\
                \multicolumn{1}{c}{Source} & \multicolumn{3}{c}{correction$\rm \left(10^9\ M_{\odot}\right)$} & 
\multicolumn{3}{c}{correction$\rm \left(10^9\ M_{\odot}\right)$} & $T_d^a$ & $Z$ & \\
  & CO & \CI~ & Dust & CO & \CI~ & Dust & $\left(K\right)$ & $\left(Z_{\odot}\right)$ & References \\ 
                \hline
                SPT0113-46 & \phantom{0}0.79\,$\pm$\,0.36 & 1.09\,$\pm$\,0.30 & 0.58\,$\pm$\,0.21 & \phantom{0}1.50\,$\pm$\,0.69 & 1.70\,$\pm$\,0.47 &
1.01\,$\pm$\,0.37 & 19.8 & \phantom{0}0.99\,$\pm$\,0.24 & 2,3,4\\
                ALESS073.1 & \phantom{0}1.20\,$\pm$\,0.60 & ... & 2.42\,$\pm$\,0.85 & \phantom{0}2.53\,$\pm$\,1.26 & ... & 3.93\,$\pm$\,1.38 & & 
\phantom{0}2.5\phantom{0}\,$\pm$\,0.9\phantom{0} & 1,5 \\ 
                SPT0345-47 & \phantom{0}2.58\,$\pm$\,1.18 & ... & 2.14\,$\pm$\,0.79 & \phantom{0}4.95\,$\pm$\,2.28 & ... & 3.86\,$\pm$\,1.42 & 32.7 
& \phantom{0}6.1\phantom{0}\,$\pm$\,2.0\phantom{0} & 2,3,4 \\
                SPT0418-47 & \phantom{0}0.44\,$\pm$\,0.21 & 0.58\,$\pm$\,0.18 & 0.22\,$\pm$\,0.08 & \phantom{0}0.81\,$\pm$\,0.34 & 0.91\,$\pm$\,0.28 &
0.39\,$\pm$\,0.15 & 18.4 &\phantom{0}0.73\,$\pm$\,0.20 & 2,3,4\\
                SPT0441-46 & \phantom{0}0.91\,$\pm$\,0.43 & 1.21\,$\pm$\,0.55 & 0.92\,$\pm$\,0.34 & \phantom{0}1.84\,$\pm$\,0.80 & 2.00\,$\pm$\,0.91 &
1.76\,$\pm$\,0.66 & 21.9 &\phantom{0}2.7\phantom{0}\,$\pm$\,0.8\phantom{0} & 2,3,4\\
                DLA0817g & \phantom{0}1.05\,$\pm$\,0.55 & ... & ... & \phantom{0}1.96\,$\pm$\,1.03 & ... & ... &  & \phantom{0}1.5\phantom{0}\,$\pm$\,0.9\phantom{0} & 6 \\
                SDP81 & \phantom{0}1.61\,$\pm$\,0.69 & ... & 1.39\,$\pm$\,0.48 & \phantom{0}2.53\,$\pm$\,1.15 & ... & 1.76\,$\pm$\,0.61 & 
& \phantom{0}3.3\phantom{0}\,$\pm$\,1.1\phantom{0} & 1,7 \\
                AZTEC/C159 & \phantom{0}1.31\,$\pm$\,0.63 & ... & ... & \phantom{0}3.80\,$\pm$\,1.84 & ... & ... &  & \phantom{0}1.5\phantom{0}\,$\pm$\,1.2\phantom{0} & 8\\
                J1000+0234& ... & ... & 2.38\,$\pm$\,0.85 & ... & ... & 3.60\,$\pm$\,1.29 & &\phantom{0}0.94\,$\pm$\,0.35 & 1 \\
                GN20 & 10.40\,$\pm$\,5.54 & ... & 5.47\,$\pm$\,2.03 & 18.63\,$\pm$\,9.89 & ... & 7.73\,$\pm$\,2.87 & & \phantom{0}0.96\,$\pm$\,0.38 & 9,10 \\
                ID141 & \phantom{0}3.68\,$\pm$\,1.61 & 3.91\,$\pm$\,1.08 & 4.67\,$\pm$\,1.51 & \phantom{0}7.59\,$\pm$\,3.33 & 5.77\,$\pm$\,2.52 &
6.95\,$\pm$\,2.25 & &\phantom{0}4.9\phantom{0}\,$\pm$\,1.9\phantom{0} & 11 \\
                SPT2132-58 & \phantom{0}1.99\,$\pm$\,0.91 & 1.30\,$\pm$\,0.54 & 2.47\,$\pm$\,0.93 & \phantom{0}4.26\,$\pm$\,1.95 & 2.30\,$\pm$\,0.97 &
5.13\,$\pm$\,1.94 & 25.3  &\phantom{0}5.7\phantom{0}\,$\pm$\,1.9\phantom{0} & 2,3,4 \\
                SPT2146-55 & \phantom{0}1.79\,$\pm$\,0.85 & 3.57\,$\pm$\,1.17 & 1.87\,$\pm$\,0.70 & \phantom{0}3.68\,$\pm$\,1.72 & 6.02\,$\pm$\,1.97 &
3.66\,$\pm$\,1.36 & 26.5 & 12.9\phantom{0}\,$\pm$\,3.6\phantom{0} & 2,3,4\\
                \hline
        \end{tabular}

\begin{flushleft}
Notes: $^a$Dust temperature obtained by fitting a modified blackbody to the flux measurements at wavelengths
$\rm \ge500\ \mu m$ (see text).

References:
1: This paper;
2: \cite{reuter2020};
3: \cite{aravena2016};
4: \cite{bothwell2017};
5: \cite{coppin2010};
6: \cite{neeleman2020};
7: \cite{dye2015};
8: \cite{jimenez2018};
9: \cite{hodge2012};
10: \cite{daddi2009};
11: \cite{dye2022}.
\end{flushleft}
\end{table*}

\subsection{Results}

\begin{figure*}
        \includegraphics[width=14cm]{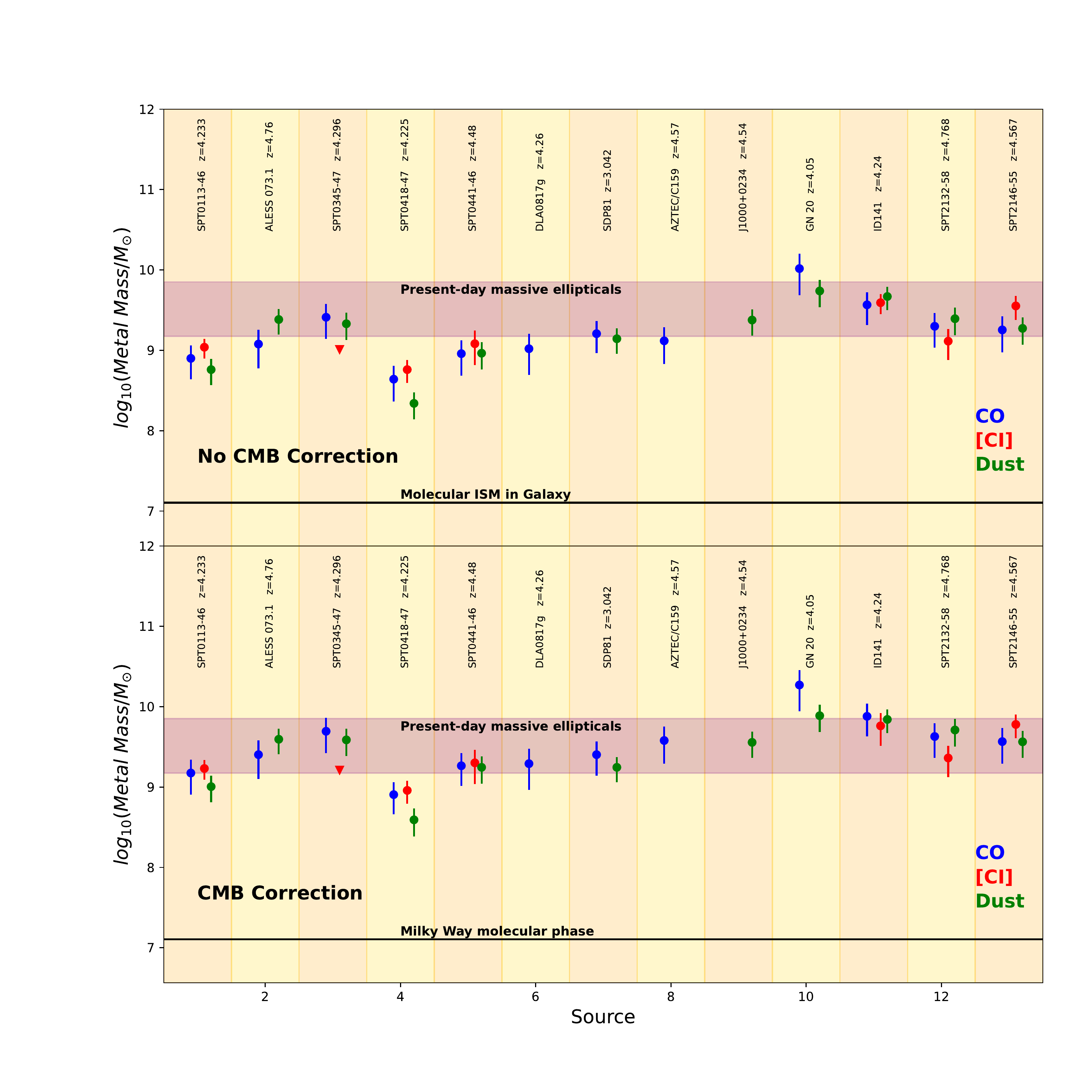}
    \caption{Estimates of the mass of metals in the molecular gas in the 13 DSFGs
estimated from CO lines (blue), the \CI~ $1-0$ line (red) and dust continuum emission
(green). The upper panel shows the estimates without a correction for the effect of the
CMB, the lower panel the estimates after
this correction has been made. The horizontal line shows an estimate of the mass of metals
in the molecular gas in the Galaxy today (see text). The horizontal purple band shows an
estimate of the range of metal mass for the most massive early-type galaxies in the
universe today (see text).}
    \label{fig:Figure1}
\end{figure*}

Table~\ref{tab:Metal Table} lists our estimates of the mass of metals using the three independent tracers, with and without 
a correction for the CMB. The results are shown in Figure~\ref{fig:Figure1}. The figure shows that there is very good agreement between the masses calculated using the different tracers.
For nine out of 10 galaxies, the different mass measurements
are consistent within the errors. The only exception is SPT0345-47, for which the estimates from
CO and dust are in very good agreement, but both are higher than the 3$\sigma$ upper limit from the \CI~ $1-0$ observation.

The line at the bottom of each panel shows an estimate of the mass of metals
in the molecular phase of the ISM in the Galaxy ($\rm 1.3\times10^7\ M_{\odot}$), which we 
have calculated from an estimate of the mass of molecular gas
in the Galaxy of $\rm \simeq 1.1 \times 10^9\ M_{\odot}$ \citep{yin2009} and estimates of the gas-to-dust ratio
(167.2) and of the fraction of metals in dust (0.52) from \cite{romanduval}. The mass of metals in the ISM in these high-redshift
galaxies is therefore $\simeq$100-1000 times greater than the mass of metals in the molecular gas in the Galaxy today.

Since it seems likely that the DSFGs are the ancestors of present-day massive
early-type galaxies (\S1), it is interesting to compare the masses of metals for the
two populations.
A useful low-redshift benchmark is the
{\it Herschel} Reference Survey (HRS), which is a volume-limited sample of galaxies
that includes the Virgo Cluster \citep{boselli2010,eales2017}. 
Since most of the metals in an early-type galaxy are contained in the stars, we have estimated
the metal mass for the HRS early-type galaxies from estimates of their stellar masses \citep{devis2017}
and a stellar metal abundance of 
${\rm log_{10}}(Z/Z_{\odot}) = 0.3$, appropriate for the galaxies with the highest
masses \citep{gallazzi2005}. In the figure, the
purple horizontal band shows the range of metal mass
for the HRS early-type galaxies, from the one with the largest stellar mass down to
the one with the tenth largest stellar mass. When the correction for the CMB is included (lower panel),
the DSFGs contain a mass of metals very similar to that in their probable descendants
in the universe today.

\subsection{Limitations of the Method}

We have used the calibration factors from \cite{dunne2022}, but their values 
are actually very similar
to the other recent best estimates in the literature. Our value
for $\rm \alpha_{\rm CO}$ - 
$\rm 4.0\ M_{\odot}\ (K\ km\ s^{-1}\ pc^2)^{-1}$) is very similar to the value adopted in the latest
big review article on the subject: $\rm \alpha_{\rm CO} = 4.36\pm0.9\ M_{\odot}\ (K\ km\ s^{-1}\ pc^2)^{-1}$ \citep{tacconi2020}.
It is lower than the value of $\rm \alpha_{\rm CO} = 6.5\ M_{\odot}\ (K\ km\ s^{-1}\ pc^2)^{-1}$ adopted by Scoville and
collaborators \citep{scoville2016} but adopting this value would make the metal masses of the DSFGs even higher.
Our value for the calibration factor for the dust emission ($\rm 1/\alpha_{850 \mu m} = 6.9 \times 10^{12}\ W\ Hz^{-1}\ M_{\odot}^{-1}$)
is very similar to the value adopted in \cite{scoville2016} ($\rm 1/\alpha_{850 \mu m} = 6.2 
\times 10^{12}\ W\ Hz^{-1}\ M_{\odot}^{-1}$). Our value for the calibration factor for atomic carbon
($\rm \alpha_{CI} = 17.0\ M_{\odot}\ (K\ km\ s^{-1}\ pc^2)^{-1}$) is also similar to the value found in the most
comparable large study \citep{heintz2020}: $\rm \alpha_{CI} = 21.4^{+13}_{-8}\ M_{\odot}\ (K\ km\ s^{-1}\ pc^2)^{-1}$.

Although our method does not require an assumption about the metal
abundance in the galaxy, it still does rely on the same assumption
as
the widely-used method for estimating the mass of molecular
gas: that the physical and chemical properties of the
ISM on which the calibration factors depend are the same
as in the Galaxy. 

\section{What do the dynamical masses tell us?}

\subsection{Estimates of the Masses}

The evidence that the gas in the DSFGs lies in a cold rotating disk
is two-fold.
First, the DSFGs all have velocity profiles (velocity versus radius) that are very similar to the velocity profiles 
of rotating disk galaxies in the universe today. Second, the galaxies mostly have ratios of $\rm v_{\rm rot}/\sigma > 10$, in which 
$\rm v_{\rm rot}$ is the rotational velocity of the gas and $\sigma$ its velocity dispersion. Such a high ratio suggests that 
gravity in the galaxy is being balanced by centripetal force rather than by the internal pressure provided by the 
spread in stellar velocities. All 13 galaxies have published estimates of their dynamical masses, which are listed in 
Table~\ref{tab: Dynamical Mass Table}. In all cases, these have been obtained using modelling packages such as  3DBarolo \citep{Teodoro2015}, 
which make it possible to vary the velocities in concentric rings around the
centre of the galaxy, and also the position of the centre and the inclination and position angle 
of the disk, until the model provides a good fit to the spectroscopic data. 

For six of the 13 galaxies, the authors of the mass estimate have made an explicit correction for the effect of the stellar 
pressure, the `asymmetric drift’. In the other studies, the authors have simply assumed 
it is negligible because of the high ratio of $\rm v_{\rm rot}/\sigma$. For each of the galaxies, 
we have used the data in the original papers to estimate the relative contributions of the asymmetric drift 
and the centripetal force, and where necessary made a correction to the published masses (Appendix B).
For all but two galaxies the correction to the mass is $\leq$10\,per\,cent.

The original papers sometimes list separate estimates of the baryonic and non-baryonic masses (where they have done this we list
the former in Table~\ref{tab: Dynamical Mass Table}).
In order to check that there are no systematic errors in the original analysis, we
we have made our own estimates of the total dynamical masses from the
velocity profiles and
the estimates of the galaxy inclination in the original papers.

We made our estimates of the mass from the equation:

\begin{equation}
M_{\rm tot} = \frac{v^2 r}{G {\rm sin}(i)^2}
\end{equation}

\noindent in which $v$ is the rotational velocity of the gas at a distance $r$ from the centre of the
galaxy and $i$ is the estimate of the inclination of the disk. We have made a correction to these
estimates for the effect of asymmetric drift (Appendix B). 
Equation 3 gives the correct mass interior to 
$r$ if the density distribution is spherically symmetric. If the density is 
not spherically symmetric our mass estimates will be too high \citep{Walter} - 
by roughly 40\,per\,cent if the mass is distributed in a razor-thin exponential disk\footnote{Bovy et al. 2022, {\it Dynamics
and Astrophysics of Galaxies}, https://galaxiesbook.org/index.html, accessed 7/4/2022}.
We calculated errors on the mass estimates
by combining the errors in velocity and inclination. 
Table~\ref{tab: Dynamical Mass Table} lists our mass estimates and the distance from the centre of the DSFG at which
we made the estimate.

The published mass estimates and own estimates are in good agreement for 10 out of the 13
galaxies, which agrees with the evidence that non-baryonic matter is neglible in the
central regions of high-redshift galaxies \citep{Genzel}.
For the three galaxies
where there is a significant discrepancy, our mass estimate is higher than the published
mass by a factor of 1.6 for 
ALESS073.1, by a factor of 2.1 for SPT0418-47 and by a factor of 1.5 for SPT0441-46.

\begin{table*}
       \caption{Dynamical Masses}
        \label{tab: Dynamical Mass Table}
        \begin{tabular}{lccccl} % four columns, alignment for each
                \hline
                 & Mass from & Our Mass & \multirow{2}{*}{Radius} & Correction for &\\
                 \multicolumn{1}{c}{Source}&  literature & estimate &  & asymmetric drift &  \multicolumn{1}{c}{Reference}\\
                 & ($\rm 10^{11}\ M_{\odot}$) & ($\rm 10^{11}\ M_{\odot}$) & (kpc) & (percentage) & \\
                \hline
                SPT0113-46 & 1.1\phantom{0}\,$\pm$\,0.1\phantom{0} & 0.93\,$\pm$\,0.13 &  3.2 & 1.0 & Rizzo et al. (2021) \\
                ALESS073.1 & 0.55\,$\pm$\,0.13 & 0.89\,$\pm$\,0.19 & 3.5 & 5.0 & Lelli et al. (2021) \\
                SPT0345-47 & 0.40\,$\pm$\,0.03 & 0.48\,$\pm$\,0.08 & 3.0 & 8.2 & Rizzo et al. (2021) \\
                SPT0418-47 & 0.25\,$\pm$\,0.02 & 0.53\,$\pm$\,0.08 & 3.5 & 2.1 & Rizzo et al. (2020) \\
                SPT0441-46 & 0.32\,$\pm$\,0.02 & 0.48\,$\pm$\,0.09 & 2.0 & 1.4 & Rizzo et al. (2021) \\
                DLA0817g & 0.91\,$\pm$\,0.29 & 0.90\,$\pm$\,0.19 & 4.2 & 26 & Neelman et al. (2020) \\
                SDP81 & 0.38\,$\pm$\,0.06 & 0.41$\pm0.08$ & 1.5 & 9.8 & Dye et al. (2015) \\
                AZTEC/C159 & 1.4\phantom{0}\,$\pm$\,0.7\phantom{0} & 1.8\phantom{0}\,$\pm$\,1.1\phantom{0} & 3.0 & 0.2 & Fraternali et al. (2021) \\
                J1000+0234& 2.3\phantom{0}\,$\pm$\,0.3\phantom{0} & 2.7\phantom{0}\,$\pm$\,0.3\phantom{0} & 4.0 & 0.8 & Fraternali et al. (2021) \\
                GN20 & 6.4\phantom{0}\,$\pm$\,2.8\phantom{0} & 6.3\phantom{0}\,$\pm$\,1.5\phantom{0} & 7.0 & 18 & Hodge et al. (2012) \\
                ID141 & 1.0\phantom{0}\,$\pm$\,0.5\phantom{0} & 0.95\,$\pm$\,0.29 & 2.0 & 2.0 & Dye et al. (2022) \\
                SPT2132-58 & 0.39\,$\pm$\,0.04 & 0.38\,$\pm$\,0.08 & 4.0 & 1.3 & Rizzo et al. (2021) \\
                SPT2146-55 & 0.22\,$\pm$\,0.02 & 0.23\,$\pm$\,0.04 & 2.7 & 2.0 & Rizzo et al. (2021) \\
                \hline
        \end{tabular}
\end{table*}

\subsection{The Mass of the Molecular Gas Versus Dynamical Mass}

We estimated the mass of the molecular gas in each galaxy using equation (1) rather than equation (2) but otherwise followed
the procedure described in Section 3.1, which means our mass estimates
are equal to the metal mass estimates in Table~\ref{tab:Metal Table}
multiplied by the gas-to-metal ratio (86.9).
Since the dynamical masses we have measured ourselves are slightly larger than the ones in the literature,
and therefore place a less stringent upper limit on the mass of molecular
gas, to be conservative we have used our mass estimates
rather than the published ones.

Ideally, we  would compare the dynamical mass with an estimate
of the mass of molecular gas made within the same radius used to estimate
the dynamical mass. 
This is possible for the observations of the dust, for which there are high-resolution observations which
show the emission is coming from
within the radius listed in Table~\ref{tab: Dynamical Mass Table} \citep{daddi2009,dye2015,spilker2016,dye2022}).
However, most of the CO and \CI~ observations
\citep{aravena2016,bothwell2017}
do not
have enough resolution to show definitively whether the CO and \CI~ emission is coming from within the region covered by
the dynamical analysis. In the case of the CO observations, this is because of the lack of a telescope with
enough resolution and in the case of the CI this is because of the need for impractically long integration times
 with
ALMA.

Nevertheless, there
is a high-resolution map of 
\CI~ $1-0$ for one galaxy \citep{dye2022} and 
there is one galaxy that has a large enough angular size that CO 1-0 observations have enough resolution
\citep{hodge2012}. In both galaxies, the line emission from the tracer
does come from within the region covered by the dynamical analysis. 
Furthermore, detailed radiative modelling of multiple CO and \CI~lines from 24 high-redshift
DSFGs \citep{harrington2021} suggests that the emission in these lines
is typically from a region within 3 kpc of the centre of the galaxy, very similar
for most of the galaxies to the radius of the region covered by the dynamical analysis
(Table~\ref{tab: Dynamical Mass Table}).
However, we cannot say for certain that in all 13 galaxies
the \CI~ and CO emission is confined to the region covered by the dynamical analysis. 

\begin{figure*}
        \includegraphics[width=14cm]{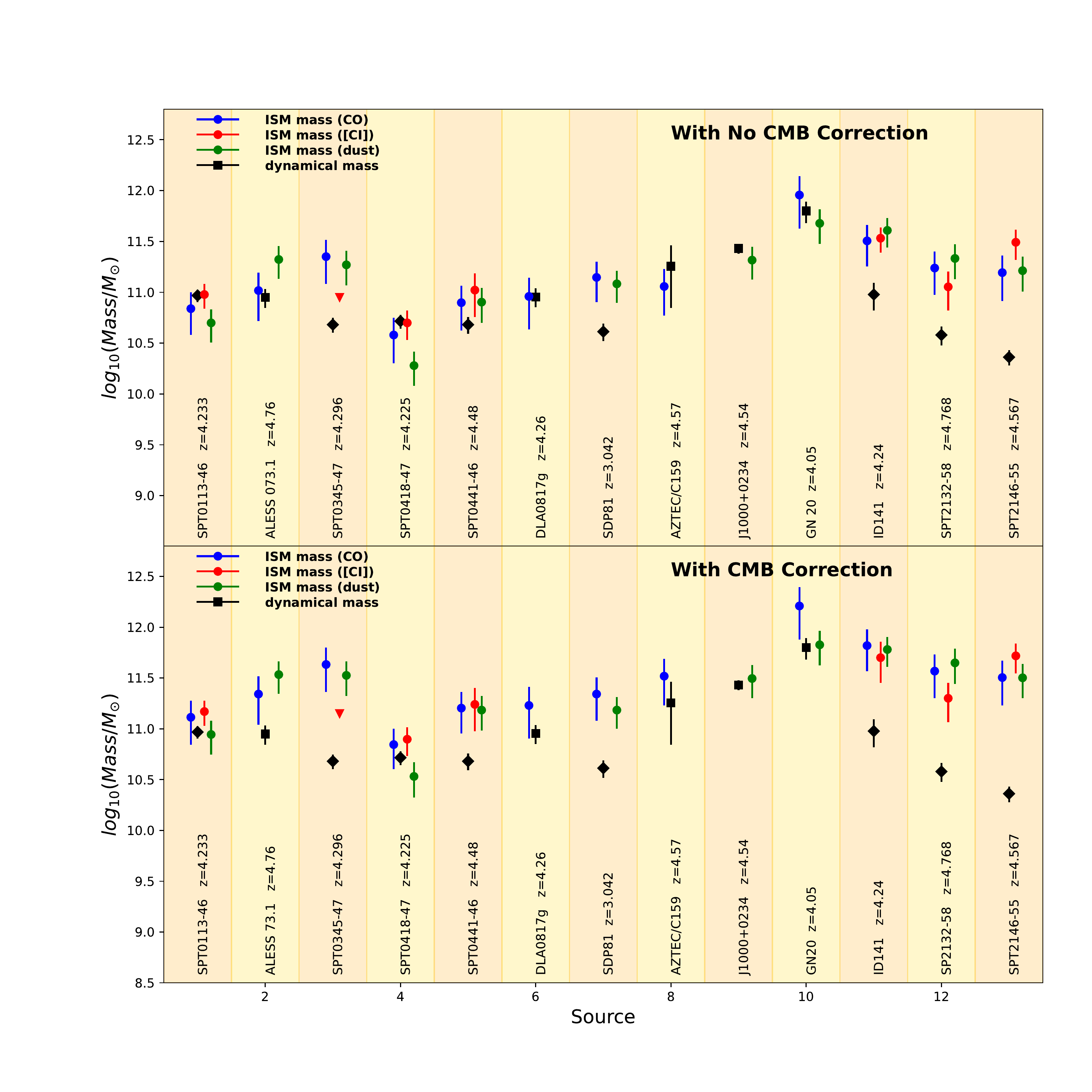}
    \caption{A comparison of the dynamical mass of each DSFG with an estimate of the mass of its gas made
using three different tracers:
carbon monoxide (blue), atomic carbon (red), dust grains (green). The 
estimate of the dynamical mass is shown by a black diamond if the source is 
gravitationally lensed and a black square otherwise. The upper panel show the 
estimates of the gas mass with no correction for the CMB,
the lower panel the estimates after  
this correction hs been made (see text).} 
    \label{fig:Figure2}
\end{figure*}

Figure~\ref{fig:Figure2} shows the comparison between the dynamical mass and the mass of
the molecular gas. The dynamical mass is a measure of the total
mass, which is the sum of the masses of
all the galaxy's components: gas, stars, and non-baryonic matter. But Figure~\ref{fig:Figure2} shows that for many of the galaxies 
the estimate {of the mass of the gas} is actually higher than the estimate of the
total mass. This paradox has been spotted many times
before when CO has been used as a tracer, but the figure shows that there is the same problem when \CI~ and dust
are used. There is, of course, the caveat that, as we noted above, for many of the galaxies
we cannot be sure that the \CI and CO emission is confined within the region covered by the dynamical
analysis, but we note that the discrepancies when dust is used as a tracer are just as large as for the other two
tracers. Moreover, the galaxy for which there is a high-resolution map in \CI~1-0, ID141 \citep{dye2022}, is one of the
best examples of this paradox.

\subsection{Possible Solutions  of the Paradox}

This paradox has been noted many times before in papers in which CO has been used to estimate
the gas mass.
Our results show that the same problem is present when carbon atoms and dust grains are used. Unless the assumptions used in the dynamical estimates are 
completely wrong, or the emission from the tracer is mostly from outside
the region covered by the kinematic observations (\S 4.2), the only way to
make the gas masses consistent with the dynamical masses is to
reduce the
values of the three calibration factors.

A crucial point to note when thinking about this issue is that the
good agreement between the gas mass estimates made with the different tracers
implies that if the calibration factors are lower than the Galactic values, all three
must be lower by a similar factor. This conclusion is supported by
the analysis in \citet{dunne2022} of 407
galaxies in the redshift
range $0 < z < 5$.  By examining
the ratios of the luminosity of the emission from the three tracers,
these authors concluded that there was no difference in the ratios
of the calibration factors between 
normal galaxies and DSFGs
and between
DSFGs
at $z < 3$ and $z > 3$.

There are two possible explanations of why the calibration factors
for the DSFGs might be lower than the Galactic values.
The first is that the difference is caused by some differences between the
ISM in the galaxy and in the Galaxy in the 
critical properties on which the calibration factors depend.
The explanation that is commonly made for
a low value of $\rm \alpha_{CO}$, for example, is that the molecular gas
in a DSFG has a higher temperature and lower density than in the Galaxy \citep{downes1998}. In Appendix A, we consider 
the properties of the ISM 
that might lead
to a change in the calibration factor.
For each tracer, there is one plausible difference between the
ISM in a high-redshift DSFG and the Galactic ISM that might lead to a change
in the calibration factor.

In the case of CO, the obvious possibility, suggested by the high star-formation rate
in these galaxies, is that the ISM
in a high-redshift DSFG is warmer and less dense than the
Galactic ISM, which would lead to a lower value of
$\alpha_{CO}$. The original motivation for this
suggestion were some CO observations which suggested
that this is the case in two low-redshift DSFGs
\citep{downes1998}. It is worth noting, however, that the results
from the first comprehensive multi-line investigation of CO in a large sample of
high-redshift DSFGs
\cite{harrington2021} suggest that the ISM in these galaxies
is not dominated by a low-density component.
The authors of this study concluded
that much of the CO
in the high-redshift DSFGs is in warm dense gas, which they argued
would have been missed by the observations in 
the low-J CO transitions used to observe the two low-redshift DSFGs \citep{harrington2021},
leading to an estimate of the value of $\rm \alpha_{CO}$ in the earlier
study that was too low.

In the case of \CI, the most plausible possibility, suggested by the high star-formation rates
in these galaxies, is that an increase in the density of cosmic rays leads to a change
in the carbon chemistry, increasing the abundance of \CI relative to CO
\citep{bisbas2015,bisbas2021,glover2016,gong2020,dunne2022} and decreasing
the value of $\rm \alpha_{CI}$.

In the case of the dust, there are several ways that the dust grains might differ from those
in the Galaxy: in their chemical composition, structures, sizes and shapes, all of which
might change the value of the calibration factor \citep{clark2016}. The obvious
way to make the value of the calibration factor lower in a DSFG is if the sizes of the
grains there are generally smaller than in the Galaxy.

The essential problem with this explanation is that each of the three calibration factors
depends on very different properties of the ISM. Even if there was any evidence that
the critical properties of the ISM on which the calibration factors depend are different in
high-redshift galaxies, which isn't the case even for CO \citep{harrington2021}, it would be a remarkable coincidence
if they were different in such a way that the calibration factors for all three tracers
were reduced by the same amount.

The second simpler explanation is that the metal abundance in the DSFGs is
higher than in the Galaxy today. This provides a straightforward explanation of 
why the calibration factors for dust and \CI are lower by the same factor because the
emission from both is optically-thin and inversely proportional to metal abundance.
It is not a perfect explanation because the relationship
between the optically-thick CO emission and
metal abundance is more uncertain, but a compilation of the data
is at least consistent with a similar relationship to that for the other two tracers
\citep{bolatto2013} (see their Figure 9), and if so the calibration factor for the
CO would decrease in the same way as for the other two tracers.

For the rest of this paper, we will follow Occam's Razor
and assume that the second explanation is the correct one, although
we 
cannot rule out the first.
If we are wrong and
the calibration factors are lower by a factor of X because of differences in the physics/chemistry/structure of
the ISM/dust grains,
our estimates of the metal masses
in \S 3 will be too high by the same factor.

\subsection{Metal Abundances}

\begin{figure}
        \includegraphics[width=\columnwidth]{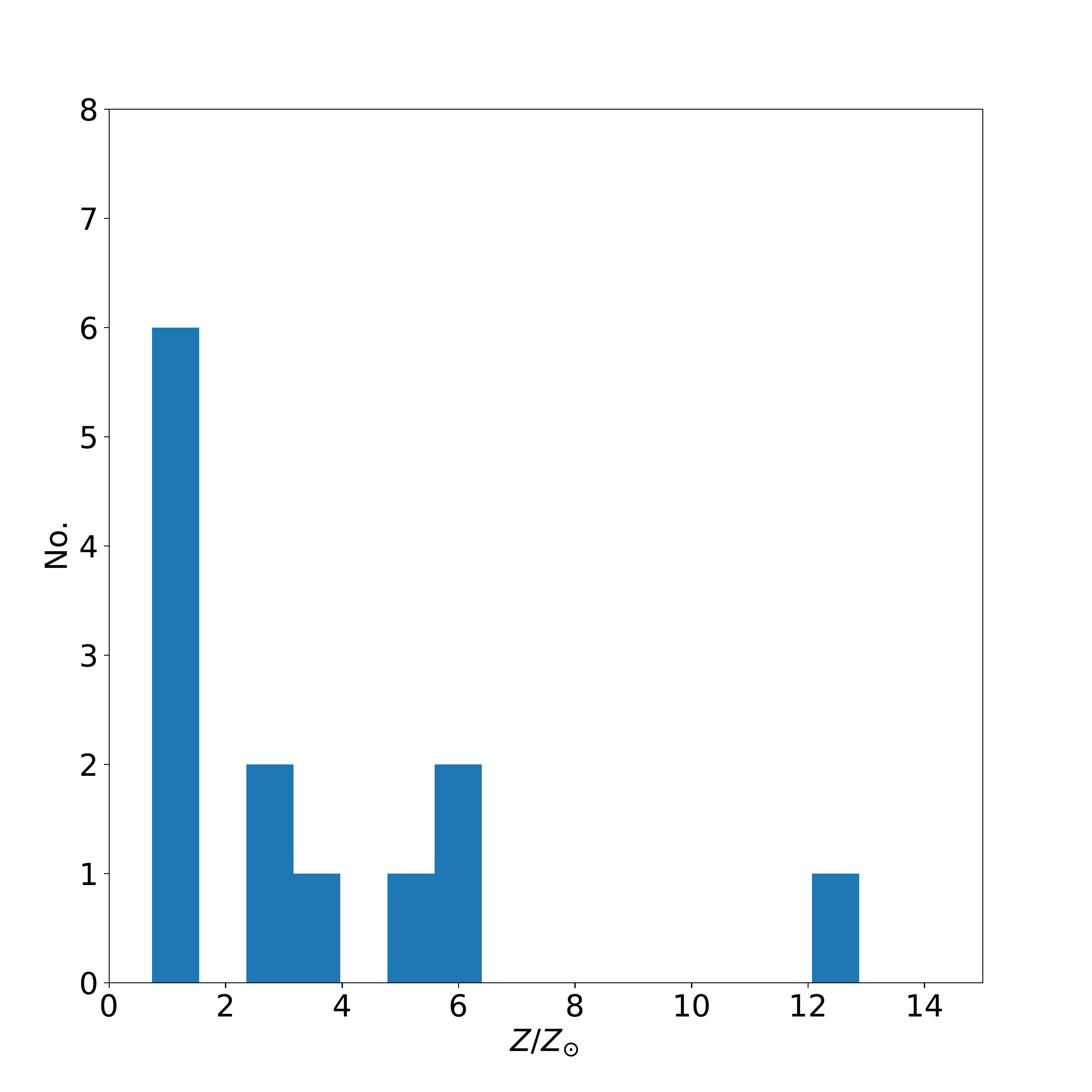}
    \caption{Estimates of the lower limit on the metal abundance for the 13 galaxies
(\S 4.4).}
    \label{fig:Figure3}
\end{figure}

We have estimated a lower limit to the metal abundance of each galaxy
from the equation:

\begin{equation} 
\frac{Z}{Z_{\odot}} > \frac{M_{\rm metals}}{M_{\rm dyn} \times 0.0142}
\end{equation}

\noindent in which $M_{\rm metals}$ is our estimate of the mass of the
metals, in which the corrections for the CMB have been
included, $M_{\rm dyn}$ is the dynamical
mass, and the number on the right-hand side is the bulk metal fraction of the Sun 
\citep{asplund2009}.
Our estimates of the limits on $Z$, which range from 0.9$Z_{\odot}$ to 12.9$Z_{\odot}$,  
are listed in Table~\ref{tab:Metal Table} and a histogram of the
values is shown in Figure~\ref{fig:Figure3}.

There are estimates of the metal abundances in the literature for three of our targets. The previous estimates and our lower limits are consistent for two galaxies but not for the third.
For ALESS073.1 the previous estimate, from far-infrared spectral lines, is that $Z/Z_{\odot}$ is
in the range 0.6-3 \citep{breuck2014}, which is consistent with our lower limit
of $2.5\pm0.9$. For SPT0418$-$47, the previous estimates, from far-infrared line measurements
($0.3 < Z/Z_{\odot} < 1.3$, \citet{breuck2019}) and from JWST spectroscopy ($Z/Z_{\odot} \simeq 1.6$, \citet{peng2023}, are consistent with our lower limit ($Z/Z_{\odot} >0.73\pm0.2$.
The inconsistency is for SDP81. There are two estimates from far-infrared spectral lines:
$Z/Z_{\odot}< 2$ \citep{rigo2018} and $Z/Z_{\odot} \simeq 0.5$ \citep{rybak2023}.
The second at least is inconsistent with our lower limit: $Z/Z_{\odot} > 3.3\pm1.1$.

\section{Metals inside and outside 
galaxies - chemical evolution models}

Such high metal 
masses and metal abundances at such early times 
are perhaps surprising because there would have been much less time to make the 
metals. We have used
two chemical-evolution models to investigate 
whether it is possible to make such large masses of metals so quickly.

The first is the
widely used gas-regulation model \citep{lilly2013,peng2014}.
In this model, there is a flow of gas from 
the intergalactic medium into the galaxy, and 
there is also an outflow 
with a rate proportional to
the galaxy's star-formation rate, 
which leads eventually to the mass of the ISM reaching an equilibrium value. The mass 
of metals that is produced in this model depends 
on the strength of the outflow
and the form of the stellar initial mass function (IMF).

The gas-regulation model is based on the 
`instantaneous recycling approximation’, 
in 
which it is assumed that all the 
metals produced by a newly 
formed population of stars are made the moment the stars are born 
and immediately released into the ISM. This assumption is likely to
be a poor one for high-redshift DSFGs
because 
of their very short gas depletion times
 \citep{dye2015,dye2022}. We have therefore also included the predictions of a model 
that does not include this assumption, 
which we constructed to model the evolution of one of 
the galaxies in our sample \citep{dye2022}. This model also 
includes more realistic inflows and outflows and takes account 
of the dependence of stellar yields on metal abundance.

\subsection{The Gas-Regulation Model}

In this model, there is an outflow
with a rate equal to a constant ($\Lambda$) times the 
galaxy's star-formation rate, which leads to the 
mass of the ISM eventually 
reaching an equilibrium value (hence the 
term `gas regulation’). We use the analytic formalism 
of \cite{peng2014}, which makes it possible to follow the evolution
of a galaxy from $t=0$ to $t >> \tau_{\rm eq}$, 
when the galaxy asymptotically approaches an 
equilibrium state. 
In this equilibrium state, the 
gas mass, the metal abundance in the gas and the mass of metals in the gas
are 
constants, although the mass of stars
continues to rise as new stars are born.

The equilibrium time, $\tau_{\rm eq}$ is given by

\begin{equation}
\tau_{\rm eq} = \frac{\tau_{\rm dep}}{1 - R + \Lambda} = 
\frac{1}{\epsilon(1 -R + \Lambda)}
\end{equation}

\noindent in which
$\tau_{\rm dep}$ is the depletion time (the 
ratio of the gas mass to the star-formation rate), $\epsilon$
is the inverse of this (the star-formation efficiency), 
and $R$ is the fraction of the mass of a 
cohort of newly formed stars that is eventually returned 
to the ISM, which in the instantaneous recycling approximation 
happens immediately. We have expressed the time dependence
of our models in units of the equilibrium time so they apply to all
galaxies and do not
vary with the values of $\epsilon$, $R$ and $\Lambda$ in individual galaxies.

We have used the following equations taken directly from \cite{peng2014}:

\begin{equation}
M_{\rm gas} = \frac{\Phi}{\epsilon (1 - R + \Lambda)} \left(1 - e^{- {\dfrac{t}{\tau_{\rm eq}}}}\right)
\end{equation}

\begin{equation}
M_{\rm star} = \frac{\Phi}{\epsilon} \frac{1-R}{(1-R+\Lambda)^2} 
\left[\frac{t}{\tau_{\rm eq}} - \left(1 - e^{- \dfrac{t}{\tau_{\rm eq}}} \right) \right]
\end{equation}

\begin{equation}
Z_{\rm gas} = \frac{y}{1 - R + \Lambda} \left(1 - e^{-\dfrac{t}{\tau_{\rm eq}}}\right)
\left[ 1 - e^{-\left(\frac{t} {\tau_{\rm eq}(1 - e^{-\frac{t}{\tau_{\rm eq}}} ) }\right)}\right]
\end{equation}

\noindent In these equations, $y$ is the yield, the mass of metals produced
by one solar mass of stars, and $\Phi$ is the rate at which gas is flowing
onto the galaxy. We can combine
equations (6) and (8) to get an estimate of the total mass of metals
in the ISM:

\begin{equation}
M_{\rm metals} = Z_{\rm gas} M_{\rm gas}
\end{equation}

\noindent We then combine equations (6), (7) and (9) to calculate the ratio
of the metal mass to the total mass ($M_{\rm gas} + M_{\rm star}$). Neither this
ratio nor the metal abundance of the ISM (equation 8) depend
on the values of $\Phi$ or $\epsilon$ but they do depend on the
values of $R$, $y$ and $\Lambda$. The first two of these
depend on the choice of the IMF.

\begin{figure*}
        \includegraphics[width=\textwidth]{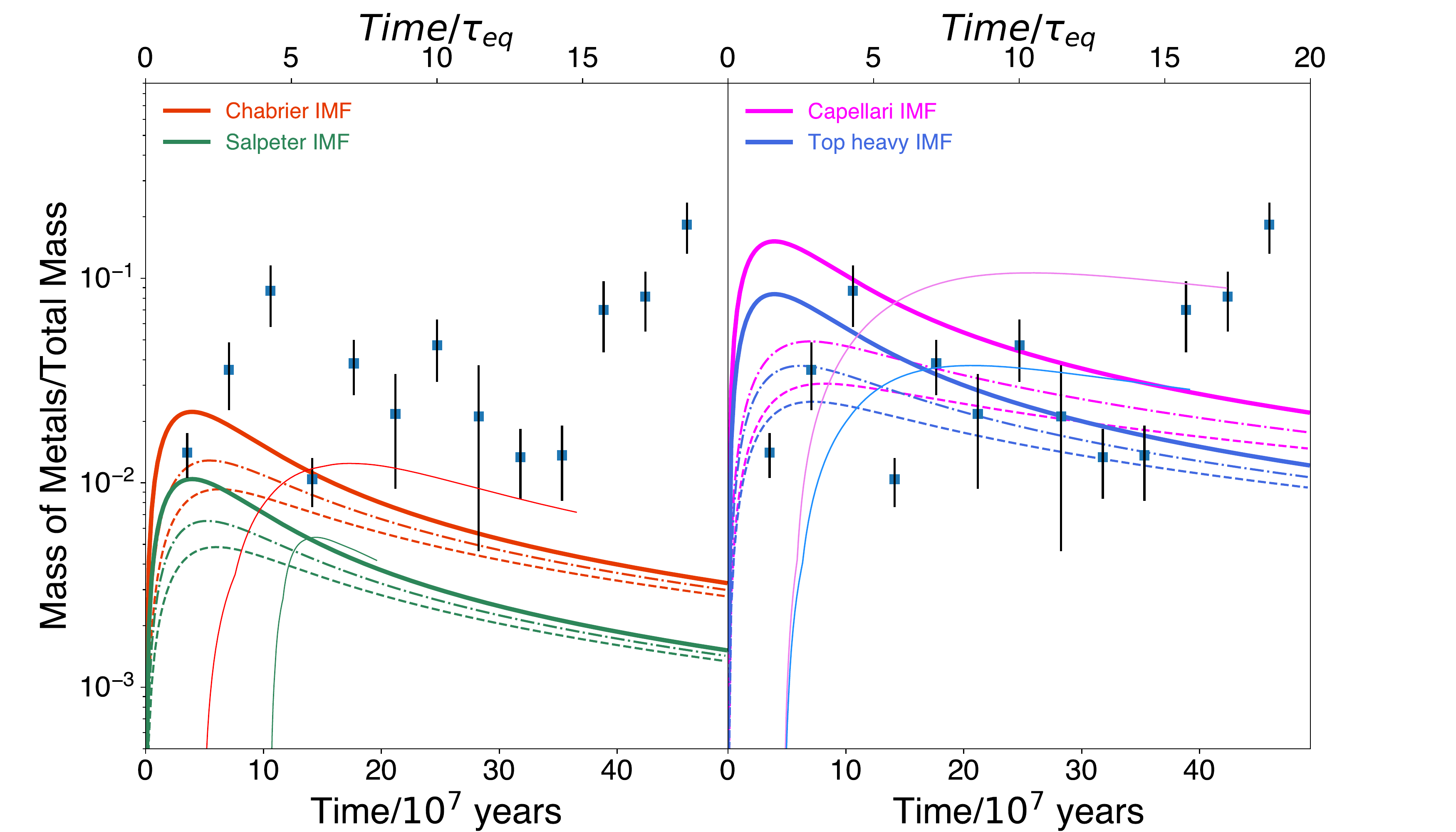}
    \caption{The ratio of the mass of ISM metals to the total
galaxy mass (gas plus stars) plotted against time. 
The squares show our estimates for the 13 DSFGs, which, since we don't know
their ages, are plotted
at arbitrary positions along the time
axis. The thick lines show the predictions of the 
gas-regulation model, in which metals are instantaneously ejected 
into the ISM. The thin lines show the predictions of a model that includes
the
delay in metal production that arises from the lifetimes of the stars. 
The left panel shows the predictions for two standard IMFs (Chabrier and Salpeter) 
and the right panel for two IMFs with a larger proportion of 
high-mass stars \citep{cappellari2012,zhang2018} (key 
at top of each panel). The continuous, dot-dashed and dashed 
lines show the predictions for different outflow 
parameters: $\Lambda=0,1,2$, respectively. 
Time is expressed in units of the equilibrium time used in the gas-regulation
model along the top of the figure and in years along the bottom of the
figure (see text).
}
    \label{fig:Figure4}
\end{figure*}

\subsection{A Bespoke  Model}

The instantaneous recycling assumption 
may be particularly poor for high-redshift DSFGs because
of their very short gas-depletion and dynamical times
\citep{dye2015,dye2022}. We have therefore also 
used a chemical evolution model we developed
for the z=4.24 DSFG ID141, 
which does not include this assumption and which incorporates
the lifetimes of the stars that make the metals \citep{dye2022}.
We refer the reader to our earlier paper for more details of the
model but, in brief, the galaxy starts as a cloud of gas with no
heavy elements, with the gas then being converted into stars using
an assumed IMF and star-formation history.
Outflows of gas and metals are based on 
prescriptions of feedback from stars and active galactic 
nuclei, and the model includes inflows via accretion from the 
cosmic web. Stars eject gas, metals and dust in each 
generation based on prescriptions for the 
stellar yield of low-mass \citep{vandenhoek1997} and 
high-mass stars \citep{maeder1992} and for 
the 
remnant mass function \citep{devis2021}. 

\subsection{Results}

We have considered four different 
forms for the IMF. The first two are the
commonly used
Chabrier and 
Salpeter IMFs. The other two are top-heavy IMFs 
in which there is a higher proportion of 
high-mass stars. One of 
these was proposed to explain the dynamics of 
low-redshift elliptical galaxies \citep{cappellari2012}, the 
other to explain the observations of isotopologues 
of carbon monoxide in four DSFGs \citep{zhang2018}. We 
have estimated the values of the return fraction, 
$R$, and the yield, $y$, for each IMF from 
a compendium of stellar yields \citep{nomoto2013}. 
The values 
are listed in Table~\ref{tab:returns}.  

\begin{figure}
        \includegraphics[width=1.1\columnwidth]{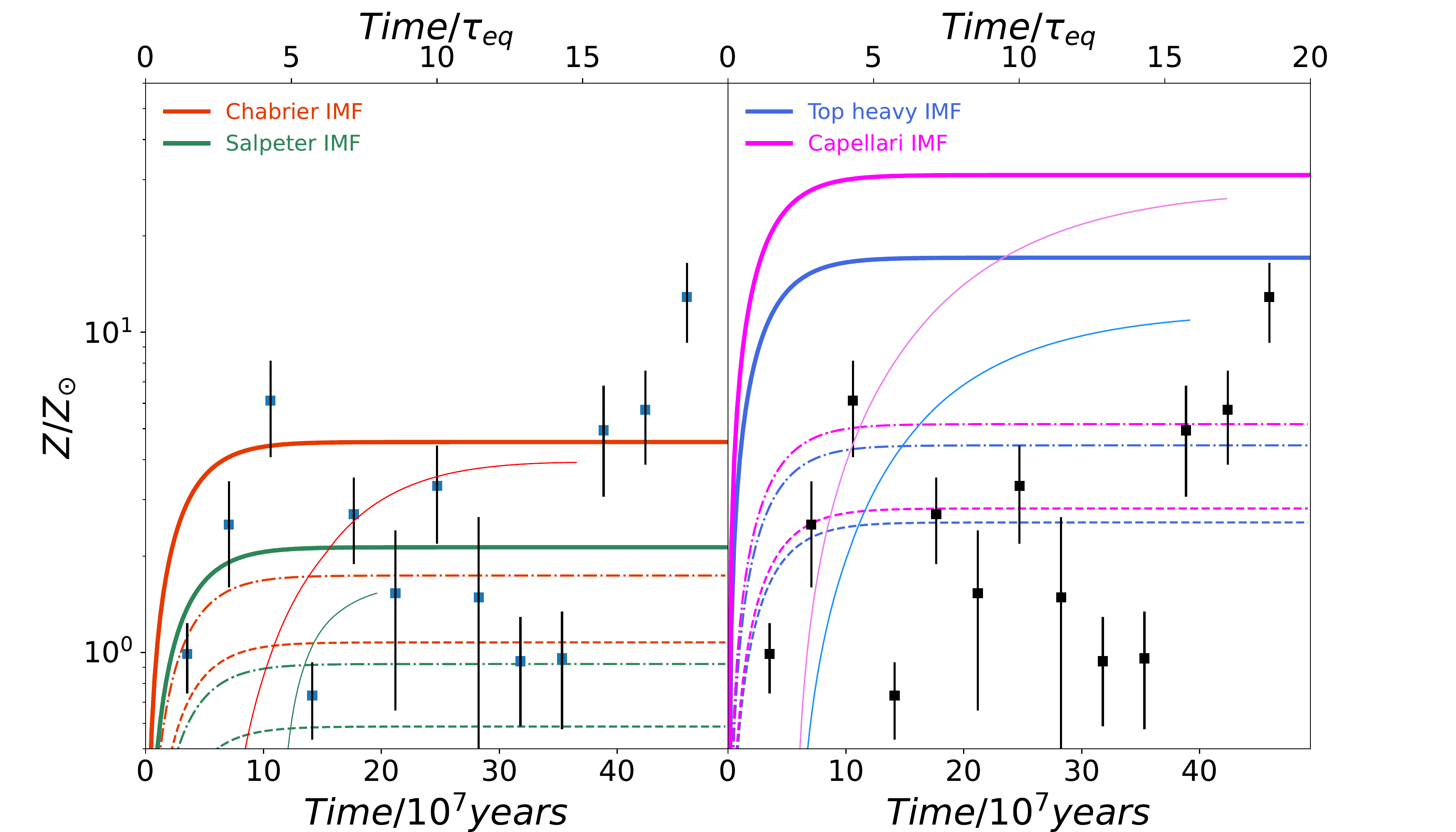}
        \centering
    \caption{Metal abundance of the ISM
plotted against cosmic time. 
The squares show our estimates of this for the 13 DSFGs (Table 3), which,
since we don't know their ages, are
are plotted
at arbitrary positions along the time
axis. The thick lines show the predictions of the
gas-regulation model, in which metals are instantaneously ejected
into the ISM. The thin lines show the predictions of a model that includes the
delay in the metal production that arises from the lifetimes of the stars.
The left panel shows the predictions for two standard IMFs (Chabrier and Salpeter)
and the right panel for two IMFs with a larger proportion of
high-mass stars \citep{cappellari2012,zhang2018} (key
at top of each panel). The continuous, dot-dashed and dashed
lines show the predictions for different outflow
parameters: $\Lambda=0,1,2$, respectively.
Time is expressed in units of the equilibrium time used in the gas-regulation
model along the top of the figure and in years along the bottom of the
figure (see text).
}
    \label{fig:Figure5}
\end{figure}

The prediction of both models
for the ratio of metal mass to total
mass (gas plus stars) versus time are shown in 
Figure~\ref{fig:Figure4} and the predictions for metal abundance versus time in Figure~\ref{fig:Figure5}. 
Time is plotted in units of the equilibrium time, a key parameter
of the gas-regulation model, along the top of each figure.
To give an impression of the timescales, we have calculated
the value of $\tau_{\rm eq}$ in years using a Chabrier IMF and the
properties of the galaxy ID141 \citep{dye2022}, and then plotted
time in the usual units along the bottom of the figures,
although strictly this is only correct for one galaxy and one
form of the IMF. We stopped the model based on ID141 \citep{dye2022} when
it reached a stellar mass of $5 \times 10^{11}\ M_{\odot}$ when we assume
an outflow clears the galaxy of its ISM \citep{romano2017}.
The left panel in each figure shows the predictions for
the standard Chabrier and Salpter IMFs and the right panels show
the predictions for the top-heavy IMFs.

In Figure~\ref{fig:Figure4}, we have compared the predictions of the models to our estimates
of the ratio of metal mass
to total mass for the DSFGs. In making these estimates, we have assumed
that the total mass is given by the dynamical mass.
In Figure~\ref{fig:Figure5}, we have compared the model predictions
to our estimates
of the lower limits on the metal abundance, which are listed in Table 3.
Since the ages of the galaxies are unknown, we have plotted
our estimates for the DSFGs in both figures at
arbitrary points along the time axis.
Note that our estimates for the DSFGs in the two figures have a simple
scaling. The difference in the two figures is that in Figure 4 we
are using the models to predict the ratio of mass of metals in the gas to
the total mass (gas plus stars) and in Figure 5 we are using the models
to predict the metal abundance in the gas.

Figure~\ref{fig:Figure4} shows that the predictions of  both 
models are very similar, with the only difference 
being, as expected, that the ratio of metal mass 
to total mass peaks later for the model that includes 
a delay from stellar lifetimes. Nevertheless, 
even in this model, the peak is reached very early, 
only $\simeq2 \times 10^8$ years after the 
galaxy begins to form, much less than the time 
since the big bang at these redshifts. 

The left-hand panel of Figure~\ref{fig:Figure4} shows that the model predictions for metal mass over total mass for
the two standard forms of the IMF are too low compared
with our estimates of this ratio
for the DSFGs.
In the case of metal abundance,
the left-hand panel of Figure~\ref{fig:Figure5} shows that it is possible to
reproduce our estimates of metal abundance (actually lower limits
on metal abundance) for a standard IMF and
a closed-box model, although not so well if there is an outflow. A closed-box model, however, seems implausible given the evidence
for outflows from star-forming galaxies in the high-redshift
universe \citep{spilker2018,jones2019,ginolfi2020,veilleux2020}.

The right-hand panels in both figures show that with a top-heavy IMF
the model predictions agree much better with our estimates.
We note that although we don't know the ages of these
galaxies, it seems likely that we are seeing them at an
epoch at which the dust mass, and therefore the mass of metals
in the ISM, was at close to its maximum value, simply because
of the strong selection bias towards high dust masses for
the galaxies discovered in a submillimetre survey.

The curves in both figures for both sets of models are sensitive to
the choice of input yields and the chosen remnant mass function
(which in turn affects the return fraction $R$). Different stellar
yield tables \citep{limongi2018,karakas2018}, 
for example, can reduce the yield, $y$, and hence
the metal abundance, $Z_{\rm gas}$, by 20-30\,per\,cent for the 
Cappellari IMF \citep{cappellari2012}.
This is not enough to affect the conclusions above.

\begin{table}
        \caption{Our Assumptions About the Return Fraction and Yield$^a$}
        \label{tab:returns}
        \begin{tabular}{lcc} % four columns, alignment for each
                \hline
                \multicolumn{1}{c}{IMF} & Yield ($y$) & Return Fraction ($R$)\\
                \hline
                Salpeter & 0.023 & 0.24 \\
                Chabrier & 0.040 & 0.38 \\
                Top-heavy \citep{cappellari2012}  & 0.088 & 0.80 \\
                Top-heavy \citep{zhang2018}& 0.085 & 0.65 \\
                \hline
        \end{tabular}
$^a$We use the definition of yield in \citet{peng2014}.
\end{table}

\subsection{The Metals in the Intracluster Gas}

Our analysis shows that the DSFGs
contain very large masses of metals. There is plenty of evidence
for massive outflows of gas from high-redshift
DSFGs \citep{spilker2018,jones2019,ginolfi2020,veilleux2020}, suggesting that
a significant proportion of the metals produced in the galaxies should have
been ejected into intergalactic space.
In this section, we 
will investigate the possibility that these
ejected metals might be the solution to the
long-standing conundrum that $\simeq$75\,per\,cent of the metals 
in rich clusters of galaxies are in the intergalactic gas rather than in the galaxies
themselves 
\citep{loewenstein2013,renzini2014}. 

We have investigated this possibility by extending the gas-regulation model (\S 5.1).
On the assumption that the metal abundance in the outflowing gas is the same
as in the ISM in the galaxy, the mass of metals ejected into intergalactic
space is given by:

\begin{equation}
M_{\rm ejected\ metals} = \int^{t_{\rm final}}_0 Z_{\rm gas} \Lambda {\rm SFR} dt
\end{equation}

\noindent in which ${\rm SFR}$ is the star-formation rate, which is given by:

\begin{equation}
{\rm SFR} = \frac{\Phi}{1 - R + \Lambda} (1 - e^{-\frac{t}{\tau_{\rm eq}}})
\end{equation}

\noindent The predicted mass of metals that is incorporated in the stars within
a galaxy is given by:

\begin{equation}
M_{\rm metals,\ stars} = (1-R) \int^{t_{\rm final}}_0 Z_{\rm gas} {\rm SFR} dt
\end{equation}

\noindent The predicted metal abundance of the stars is then given by:

\begin{equation}
Z_{\rm stars} = \frac{ \int^{t_{\rm final}}_0 Z_{\rm gas} {\rm SFR} dt}{\int^{t_{\rm final}}_0 {\rm SFR} dt}\end{equation}

\begin{figure}
        \includegraphics[width=\columnwidth]{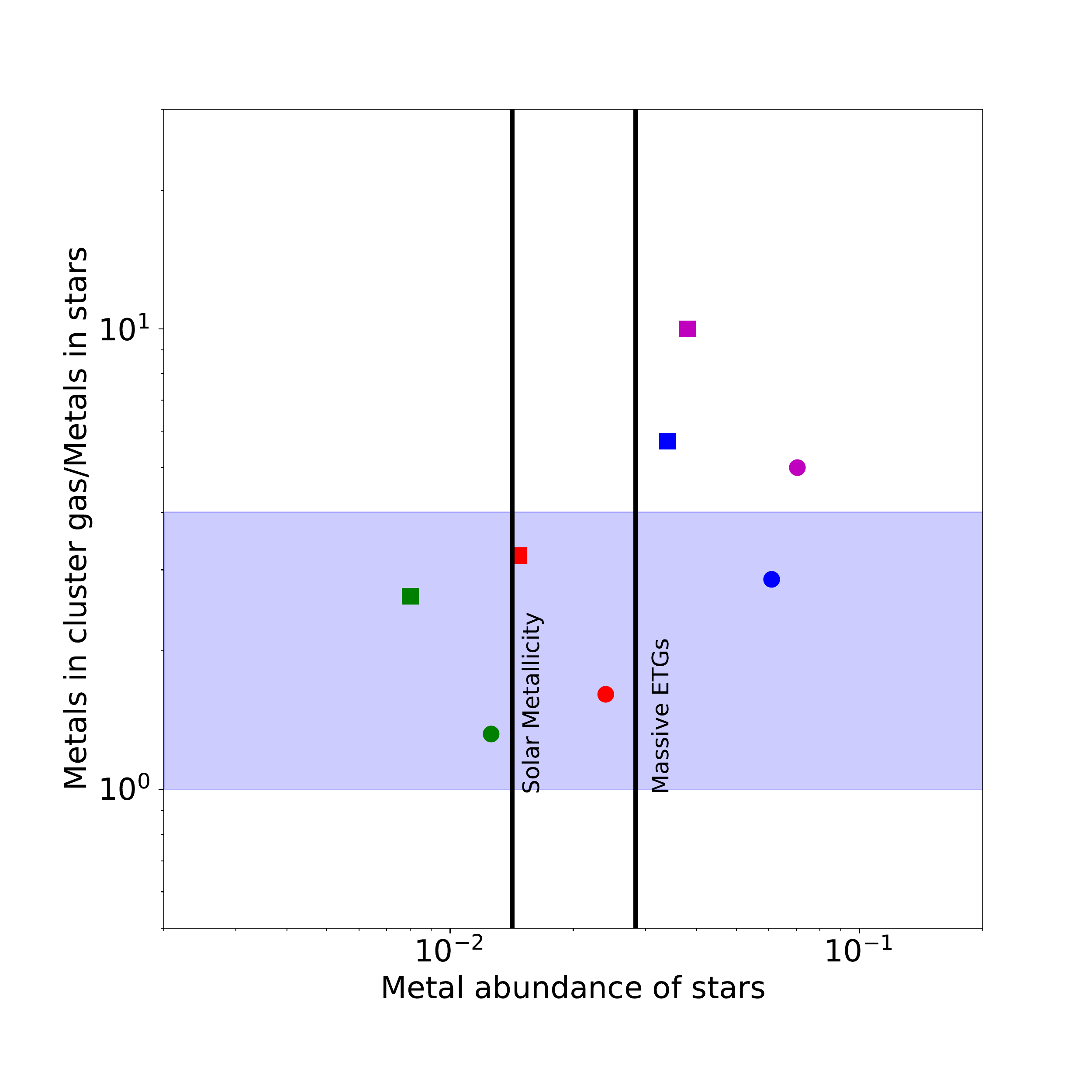}
    \caption{The ratio of the mass of metals in the gas in present-day rich clusters
    to the mass of metals in the cluster galaxies versus the metal abundance
    in the galaxies. The horizontal
    purple band shows the observed range for the mass ratio
    \citep{renzini2014}. The vertical lines show solar metal abundance \citep{asplund2009} and twice solar metal abundance, typical of
the massive early-type galaxies in cluster today.The 
coloured symbols show the prediction of eight models: blue - top-heavy IMF \citep{zhang2018}; red - Chabrier IMF; purple - top-heavy IMF \citep{cappellari2012}; green - Salpeter IMF; circles - $\Lambda = 1$; squares - $\Lambda = 2$.
}
    \label{fig:Figure6}
\end{figure}

Figure~\ref{fig:Figure6} shows a comparison with the observations of the predictions of this
model for the metal abundance in present-day cluster galaxies and for the ratio
of the mass of metals that is in the cluster gas to the mass of metals that is in the
galaxies in the cluster.
The metal abundance is given by equation (13), the ratio of the metal masses by
 equations (10) and (12).
The mass ratio is independent of
$t_{\rm final}$ and the metal abundance is
only weakly dependent on it -  we have set its value to 20 $\tau_{\rm eq}$.
We have plotted the predictions for outflows with $\Lambda$ of 1 and 2,
typical of the values for high-redshift DSFGs
\citep{ginolfi2020}, and for the four forms of the IMF.
The horizontal purple band shows the
observed range of the metal ratio for present-day rich clusters \citep{renzini2014} and the two vertical lines show solar metal abundance \citep{asplund2009} and
twice solar metal abundance, which is typical of the massive early-type 
galaxies in clusters \citep{gallazzi2005}. Several of the models predict values in roughly
the right area, showing that the outflow of metals from high-redshift DSFGs
is a quantitatively plausible solution of this puzzle.

\section{Discussion}

\subsection{The Observational Results}

We have two main observational results:
(1) the mass
of metals in the molecular gas in a high-redshift DSFG is very high, as large as the entire
metal content of a massive low-redshift early-type galaxy; (2) the metal
abundance is also very high, often well over the solar
value. In this section, we will discuss the limitations of these
results. In the following section we will discuss their implications
for galaxy evolution.

The first result seems robust. It is based on the 
estimates of the calibration factors of the CI, CO and dust
tracers made by \citet{dunne2022}, but these
are quite similar to 
the other recent estimates in the literature (\S 3.4).
The discrepncies between the ISM and dynamical mass estimates
in Figure~\ref{fig:Figure2} imply that the true calibration factors for the high-redshift DSFGs
must be lower than these
values, but as long as this
decrease is caused by
an increase in the metal abundance rather than
physical/chemical/structural changes in the ISM/dust,
our estimates
of the metal mass are unaffected (\S 4.3).

For the second result, there are two causes for concern. The first
is the accuracy of the dynamical masses. Although we
have checked the published masses by making our own estimates,
both sets 
do rely on the assumption that the gas is distributed in a cold
rotating disk. Although the velocity profiles do look remarkably like
the rotation curves of present-day disk galaxies (see, for example, 
the high-resolution
velocity profile for SDP81 in \citet{dye2015}), it is possible that these may 
yet prove to have an alternative explanation.

A more immediate concern is that we cannot be always sure the CO and CI
emission is from within the region traced by the dynamical
analysis. There have been some recent claimed detections of molecular
gas in the halos around high-redshift AGN \citep{jones2023,scholtz2023}, which, if it
was also true of the DSFGs, would reduce the conflict between the ISM and dyanamical mass
and weaken the limit on the metal abundance. 
To estimate the possible size of this effect, we 
used the results of
\citet{scholtz2023}, who used a stacking
analysis to search for extended emission in the \CI~ $2-1$ line
around
a sample of extremely red quasars. They estimated average molecular masses of
$\rm 10^{10.8\pm0.14}\ M_{\odot}$ and $\rm 10^{10.2\pm0.16}\ M_{\odot}$ for the
galaxy and halo components, respectively. If these proportions were also
true of the DSFGs, our limits on metal abundance
would be lower by $\simeq$20\%, although our estimates of the total metal masses
would be unaffected.

\subsection{Implications for Galaxy Evolution}

The metal abundance in galaxies are generally less at high redshifts, with the metal
abundance of a galaxy at $z \sim 4$ being a factor of $\simeq4$ lower than a galaxy
of the same stellar mass in the universe today \citep{cullen2019}. There is already evidence
in the literature that
the high-redshift DSFGs are an anomaly, with most estimates of the metal abundance
being at roughly solar or above \citep{breuck2014,breuck2019,wardlow2017,rigo2018,rybak2023,peng2023}.
Our results support this conclusion.

We have used chemical-evolution models to show that it is possible to
produce such high metal masses and metal abundances shortly after the
formation of a galaxy (\S 5). In these models, the
mass of metals in the ISM
is at its highest only $\rm \sim 2\times10^8\ years$ after the
galaxy begins to form, much less than the age of the universe at these redshifts.
The high values for the ratio of metal mass to total mass (Figure~\ref{fig:Figure4}) and for metal
abundance (Figure~\ref{fig:Figure5}) can only be
be reproduced by the models if the IMF has a 
higher fraction of high-mass stars than the standard IMFs.
Observations of CO isotopologues in four DSFGs also imply these galaxies
have a top-heavy IMF \citep{zhang2018}.

We used these models to show that the metals in the outflows are
a possible explanation of the metals in the gas in present-day clusters
(\S 5.4). Figure~\ref{fig:Figure6} shows that this conclusion does not require a
top-heavy IMF;
the success of the models
is due to the large 
masses of metals carried by the outflows rather than the metal
abundance,
which is the result that requires a top-heavy IMF.
Our model is, of course, very simple, and we have applied it
to a sample of extreme DSFGs, which are
likely to be the ancestors of only the most massive early-type galaxies. 
A conclusive demonstration
that the metals in galactic outflows are the source of the
metals in the intracluster gas would require the model to 
include galaxies covering the whole range of stellar masses seen in present-day clusters.

Our estimates of the metal abundance in the high-redshift DSFGs are surprisingly large,
often even larger than the values for the galaxies
that are likely to be their descendants, the most massive early-type
galaxies in the universe today, which have a metal abundance about twice solar
\citep{gallazzi2005}.
Is there a way that the subsequent evolution of the
DSFG might have reduced its initial metal abundance?
The 
evolutionary route from a DSFG, a galaxy with a small physical size but with a
huge star-formation rate, to a massive early-type galaxy in the universe
today, a galaxy with a large
physical size but a low star-formation rate,
is still very uncertain \citep{tachella2016}.
We speculate that it is the sequence of mergers that must have occurred
to increase the size of the galaxy that has 
reduced its initial metal abundance.
Since models of the stellar populations of
present-day early-type
galaxies imply that relatively few stars were formed after the
DSFG epoch
\citep{thomas2010}, it seems
likely that most of these mergers would have been dry mergers, with the
DSFG-descended galaxy always being the most massive member of the merger because
otherwise its mass would have grown too much by the present day.
Given the
strong relationship between metal abundance and stellar mass seen at all
epochs (\S 1), it therefore seems likely that each merger would have further reduced
the metal abundance.

\section{Conclusions}
 
We have estimated the mass of metals in the molecular gas in 
13 dusty star-forming galaxies (DSFGs) at $z \sim 4$ for which
previous observations have shown that the gas lies
in
a cold rotating disk. We estimated the metal masses using observations
of CO in either the $\rm J=1-0$ or $\rm J=2-1$ lines, observations of atomic carbon in the \CI~ $1-0$ line and
continuum observations of dust.
In making our estimates, we used the first mutually-consistent estimates of the calibration factors
for the three tracers
\citep{dunne2022}. There were observations of
at least two tracers 
for 10 out of 13 galaxies and observations of all three for 7 galaxies.
Our method is independent of the
metal abundance in the galaxy, in contrast to the widely used  method in which the tracer
is used to estimate the ISM mass. We obtained the following results:

\begin{itemize}

\item We obtained very similar mass estimates from the different tracers, our estimates
being 
similar to the entire metal content of a massive present-day early-type
galaxy.

\item When we estimated the mass of the ISM
rather than the mass of the metals, we found that our estimate was often higher than the estimate of the dynamical mass. This paradox has been
noticed before for CO, but the same is true when carbon atoms or dust grains
are used as the tracer.
 
\item We argue that the most likely solution of this paradox is that the metal
abundances in the high-redshift DSFGs are often higher than in the Galaxy. The alternative explanation, that the calibration factors
are lower because of differences in the physics/chemistry/structure of the
ISM/dust, seems unlikely because these differences would have to conspire to
reduce all three calibration factors
by the same amount.

\item On the
assumption that our solution
is the correct one, we estimated lower limits to the
metal abundance 
($Z/Z_{\odot}$) in the molecular gas in these galaxies of between 0.9 and 12.9. The main caveat 
is that we cannot be sure for all galaxies that the CO and \CI~ emission
is confined to the region covered by the dynamical analysis. 

\item We have used chemical evolution models to 
show that it is possible to produce such high metal masses and abundances
shortly after the formation of a galaxy as long as the stellar IMF in the galaxy
is top-heavy.

\item We used these models to show that the metals in the outflows from these galaxies can explain quantitatively
the long-standing conundrum that $\simeq$75\,per\,cent of the metals in present-day rich clusters are in the
intracluster gas rather than in the galaxies.

\item Our estimates of the metal abundance in the DSFGs are sometimes higher than the values of the metal
abundance in the galaxies today that are their probable descendants. We speculate that
the explanation is the gradual dilution of the metal content by a sequence of dry mergers.

\end{itemize}

\section*{Acknowledgements}

We thank Tim Davis for comments on an early version
of this manuscript. We thank the referee for some useful comments.
Stephen Eales and Matthew Smith
thank the Science and Technology Facilities Council for support 
(consolidated grant ST/K000926/1).

%%%%%%%%%%%%%%%%%%%%%%%%%%%%%%%%%%%%%%%%%%%%%%%%%%
\section*{Data Availability}

The ALMA data referenced in Table~\ref{tab:alma_table} is in the ALMA archive. All the other observational
data is in published papers.



\bibliographystyle{mnras}
\bibliography{metals_paper_draft2} % if your bibtex file is called example.bib

%\begin{thebibliography}{99}
%\end{thebibliography}

%%%%%%%%%%%%%%%%%%%%%%%%%%%%%%%%%%%%%%%%%%%%%%%%%%

%%%%%%%%%%%%%%%%% APPENDICES %%%%%%%%%%%%%%%%%%%%%

\appendix

\section{Differences in the ISM of a high-redshift galaxy that might lead to a change
in the calibration factors for the ISM tracers}

We briefly discuss here some of the possible differences between the ISM of a high-redshift galaxy and the ISM
of the Galaxy which might lead to a difference in the calibration factor for each of the tracers. We refer
the reader to \citep{dunne2022} for a fuller discussion. \citet{dunne2022} made the first attempt
to estimate the three calibration factors simultaneously without assuming that one was more reliable 
than the others. Their sample included 407 galaxies ranging from nearby disks to DSFGs out to $z\simeq6$.

They estimated the calibration factors by minimizing the variance in the three luminosity ratios:
$\rm L_{CO}/L_{CI}$, $\rm L_{CO}/L_{dust}$, $\rm L_{CI}/L_{dust}$. They found no evidence for any variation
of these ratios with redshift, in agreement with our conclusion that the ratios of the calibration
factors must be the same at high and low redshift. Since there is no way of measuring the mass of 
molecular gas directly, their method, like those used in all similar studies, was ultimately based on an assumption about the Galactic ISM, in their case that
the gas-to-dust ratio is the same as in the Galaxy.

We reproduce here three equations from that paper (equations 9, 10, and 11) which show how the
calibration factors depend on the properties of the ISM. We use these equations to discuss how
differences between the ISM in a high-redshift galaxy and the Galactic ISM might lead to a
change in the calibration factor.

For key equation for \CI\ is:

\begin{equation}
{\rm \alpha_{CI} = 16.8 \left[{X_{CI} \over 1.6\times10^{-5}}\right]^{-1} \left({Q_{10}\over 0.48}\right)^{-1} M_{\odot} (K\ km\ s^{-1}\ pc^2)^{-1}}
\end{equation}

\noindent in which $\rm X_{CI} = [C^0/H_2]$, the abundance ratio of carbon atoms relative to hydrogen molecules, and
$\rm Q_{10}$ is the is fraction of the carbon atoms in the J=1 state. 

The calibration factor might be different from the value assumed in this paper if either $\rm X_{CI}$ or $\rm Q_{10}$ are different from the values assumed by \citet{dunne2022}. The dependence of $\rm Q_{10}$ on the density and
temperature of the gas in none-LTE conditions is derived analytically in the appendix of \citet{papadopoulos2004}.
Figure D1 in Appendix D of {\citet{dunne2022}} shows that for a reasonable range of $\rm n$ ($\rm 300<n<10,000\ cm^{-3}$
and $\rm T_K$ ($\rm 25 < T_k < 80\ K$) $\rm Q_{10}$ does not go outside the range 0.35 to 0.53. Therefore, it
seems unlikely that changes in the density and temperature of the gas will lead to large changes in the calibration factor.
In particular, changing the value from that assumed by \citet{dunne2022} (0.48) to 0.53 will only lower the
calibration factor by a factor of $\simeq$10\%. 

The other factor, $\rm X_{CI}$ depends in a straightforward
way on the metal abundance and the fraction of carbon that is the atomic form. The first of these we don't
discuss here because this is our alternative explanation of the change in the calibration factor.
A plausible reason why the proportion of atomic carbon might be higher in a high-redshift DSFG, which would lead to a lower value of $\rm \alpha_{CI}$, is that the density of cosmic rays 
might well be higher, which would lead to a change
in the carbon chemistry and increase the abundance of \CI relative to CO
\citep{bisbas2015,bisbas2021,glover2016,gong2020,dunne2022}. We note, though, that \citet{dunne2022} do not
find any difference between the value of $\rm \alpha_{CI}$ for high-redshift DSFGs and galaxies at the same redshift
with lower star-formation rates, which would be likely to have lower densities of cosmic rays.

The key equation for dust is:

\begin{equation}
{\rm \alpha_{850} =   {1.628 \times 10^{16} \over 1.36 \kappa_H} \left( {24.5 \over T_{mw}}  \right)^{-1.4}\ W\
Hz^{-1}\ M_{mol}^{-1}  }
\end{equation}

\noindent in which $\rm T_{mw}$ is the mass-weighted dust temperature and $\rm \kappa_H$ is a constant of proportionality linking the mass of gas and the mass of dust. The latter depends in a straightforward way on the metal abundance,
the fraction of the metals in the dust grains, and the submillimetre mass-opacity coefficient (the submillimetre
opacity for a unit mass of dust). 

The explanation of a difference in the calibration factor is unlikely to
be a difference in $\rm T_{mw}$ because the values we directly estimated for six of our sample (Table 3) are
similar to the value assumed in \citet{dunne2022} (24.5 K). We don't discuss a change in the
metal abundance here because this is our preferred explanation of the change in the calibration factor.
The fraction of the metals bound up in dust grains is already high in the Galaxy (0.52, \citet{romanduval}, so it seems unlikely it would be much higher in a high-redshift DSFG. 

The most plausible way to
change this calibration factor would be if the dust grains in the DSFG are different
in some way - in their chemistry, structures, shapes or sizes - from those in the Galaxy \citep{clark2016}. The easiest way to make the calibration factor smaller would be if the dust grains are smaller than in the Galaxy, increasing the total surface of the dust 
grains for a given mass of dust.

On the large-velocity-gradient assumption, the key equation for CO is:

\begin{equation}
{\rm \alpha_{CO} = 2.65 {\sqrt{n_{H_2}} \over T_b } K_{vir}^{-1} [M_{\odot} K\ km\ s^{-1}\ pc^2]^{-1}}
\end{equation}

\noindent in which $\rm n_{H_2}$ in the average density of the molecular gas in $\rm cm^{-3}$, $\rm T_b$ is the
brightness temperature of the CO 1-0 line and $\rm K_{vir}$ describes the average
dynamic state of the gas \citep{dunne2022}. The obvious way to make this calibration
factor smaller in the DSFG if
is if
the density of the gas is lower and/or the temperature is higher, which is plausible since the intensity
of the optical/UV radiation field is likely to be higher in a high-redshift DSFG because of its
higher star-formation rate.
We note, though, that \citet{dunne2022} do not
find any difference between the value of $\rm \alpha_{CO}$ for high-redshift DSFGs and galaxies at the same redshift
with lower star-formation rates. We also note that two of the key results of the most extensive multi-line
CO observations of high-redshift DSFGs \citep{harrington2021} are (a) that the ISM in these
galaxies is dominated by warm dense gas and (b) that the calibration factor is similar to
its Galactic value.

\section{Correction for the Effect of Asymmetric Drift}

We have made a correction to the dynamical masses for the effect of stellar pressure (asymmetric drift) for
the six galaxies for which this correction was not made in the original analysis in the literature, using the
formalism given in Appendix A of \citet{Bouche2022}.
The gravitational potential is given by:

\begin{equation}
\frac{GM(<r)}{r} = v_{\rm rot}^2 + \eta \sigma^2\left(\frac{r}{r_d}\right)
\end{equation}

\noindent in which $v_{\rm rot}$ is the rotational
velocity, $\sigma$ is the velocity dispersion, $r_d$ is the scale length of the disk, and $\eta$ is a numerical constant of order unity.
The value of $\eta$ depends on the model assumed for the disk \citep{Bouche2022}.
We have assumed a value of 1 appropriate for a disk in which the disk thickness and velocity dispersion have no radial dependence.
The fractional correction to the masses for asymmetric drift is given by the ratio of the second to the first term on the righthand side of
the equation.
We have estimated this from the data in the original papers, and our estimates are listed in Table~\ref{tab: Dynamical Mass Table}.
For all but two galaxies the correction is $\leq$10\,per\,cent.
We have used these estimates to correct the masses for asymmetric drift for the
seven galaxies in the sample for which this effect was not included in the original analysis.

%%%%%%%%%%%%%%%%%%%%%%%%%%%%%%%%%%%%%%%%%%%%%%%%%%

% Don't change these lines
\bsp	% typesetting comment
\label{lastpage}
\end{document}